\documentclass[11pt,a4paper]{article}

\usepackage{float}
\usepackage{flafter}
\makeatletter

\gdef\@fpheader{Prepared for submission to JHEP}
\gdef\@journal{jhep}

\newif\ifnotoc\notocfalse
\newif\ifemailadd\emailaddfalse
\newif\iftoccontinuous\toccontinuousfalse
\newif\ifnatbibsort\natbibsorttrue

\RequirePackage{amsmath}
\RequirePackage{amssymb}
\RequirePackage{epsfig}
\RequirePackage{graphicx}
\ifnatbibsort\RequirePackage[numbers,sort&compress]{natbib}\else\RequirePackage[numbers,compress]{natbib}\fi
\RequirePackage[colorlinks=true
  ,urlcolor=blue
  ,anchorcolor=blue
  ,citecolor=blue
  ,filecolor=blue
  ,linkcolor=blue
  ,menucolor=blue
  ,pagecolor=blue
  ,linktocpage=true
  ,pdfproducer=medialab
  ,pdfa=true
]{hyperref}

\def\@subheader{\@empty}
\def\@keywords{\@empty}
\def\@abstract{\@empty}
\def\@xtum{\@empty}
\def\@dedicated{\@empty}
\def\@arxivnumber{\@empty}
\def\@collaboration{\@empty}
\def\@collaborationImg{\@empty}
\def\@proceeding{\@empty}
\def\@preprint{\@empty}

\newcommand{\subheader}[1]{\gdef\@subheader{#1}}
\newcommand{\keywords}[1]{\if!\@keywords!\gdef\@keywords{#1}\else%
\PackageWarningNoLine{\jname}{Keywords already defined.\MessageBreak Ignoring last definition.}\fi}
\renewcommand{\abstract}[1]{\gdef\@abstract{#1}}
\newcommand{\dedicated}[1]{\gdef\@dedicated{#1}}
\newcommand{\arxivnumber}[1]{\gdef\@arxivnumber{#1}}
\newcommand{\proceeding}[1]{\gdef\@proceeding{#1}}
\newcommand{\xtumfont}[1]{\textsc{#1}}
\newcommand{\correctionref}[3]{\gdef\@xtum{\xtumfont{#1} \href{#2}{#3}}}
\newcommand\jname{JHEP}

\newcommand\preprint[1]{\gdef\@preprint{\hfill #1}}

\newcommand\note[2][]{%
\if!#1!%
\stepcounter{footnote}\footnotetext{#2}%
\else%
{\renewcommand\thefootnote{#1}%
\footnotetext{#2}}%
\fi}

\newtoks\auth@toks
\renewcommand{\author}[2][]{%
  \if!#1!%
    \auth@toks=\expandafter{\the\auth@toks#2\ }%
  \else
    \auth@toks=\expandafter{\the\auth@toks#2$^{#1}$\ }%
  \fi
}

\newtoks\affil@toks\newif\ifaffil\affilfalse
\newcommand{\affiliation}[2][]{%
\affiltrue
  \if!#1!%
    \affil@toks=\expandafter{\the\affil@toks{\item[]#2}}%
  \else
    \affil@toks=\expandafter{\the\affil@toks{\item[$^{#1}$]#2}}%
  \fi
}

\newtoks\email@toks\newcounter{email@counter}%
\newcommand{\emailAdd}[1]{%
\emailaddtrue%
\ifnum\theemail@counter>0\email@toks=\expandafter{\the\email@toks, \@email{#1}}%
\else\email@toks=\expandafter{\the\email@toks\@email{#1}}%
\fi\stepcounter{email@counter}}
\newcommand{\@email}[1]{\href{mailto:#1}{\tt #1}}

\newcommand*\collaboration[1]{\gdef\@collaboration{#1}}
\newcommand*\collaborationImg[2][]{\gdef\@collaborationImg{#2}}

\newcommand\afterLogoSpace{\smallskip}
\newcommand\afterSubheaderSpace{\vskip3pt plus 2pt minus 1pt}
\newcommand\afterProceedingsSpace{\vskip21pt plus0.4fil minus15pt}
\newcommand\afterTitleSpace{\vskip23pt plus0.06fil minus13pt}
\newcommand\afterRuleSpace{\vskip23pt plus0.06fil minus13pt}
\newcommand\afterCollaborationSpace{\vskip3pt plus 2pt minus 1pt}
\newcommand\afterCollaborationImgSpace{\vskip3pt plus 2pt minus 1pt}
\newcommand\afterAuthorSpace{\vskip5pt plus4pt minus4pt}
\newcommand\afterAffiliationSpace{\vskip3pt plus3pt}
\newcommand\afterEmailSpace{\vskip16pt plus9pt minus10pt\filbreak}
\newcommand\afterXtumSpace{\par\bigskip}
\newcommand\afterAbstractSpace{\vskip16pt plus9pt minus13pt}
\newcommand\afterKeywordsSpace{\vskip16pt plus9pt minus13pt}
\newcommand\afterArxivSpace{\vskip3pt plus0.01fil minus10pt}
\newcommand\afterDedicatedSpace{\vskip0pt plus0.01fil}
\newcommand\afterTocSpace{\bigskip\medskip}
\newcommand\afterTocRuleSpace{\bigskip\bigskip}
\newlength{\affiliationsSep}
\newcommand\beforetochook{\pagestyle{myplain}\pagenumbering{roman}}

\DeclareFixedFont\trfont{OT1}{phv}{b}{sc}{11}

\renewcommand\maketitle{
\pagestyle{empty}
\thispagestyle{titlepage}
\setcounter{page}{0}
\noindent{\small\scshape\@fpheader}\@preprint\par
\afterLogoSpace
\if!\@subheader!\else\noindent{\trfont{\@subheader}}\fi
\afterSubheaderSpace
\if!\@proceeding!\else\noindent{\sc\@proceeding}\fi
\afterProceedingsSpace
{\LARGE\flushleft\sffamily\bfseries\@title\par}
\afterTitleSpace
\hrule height 1.5\p@%
\afterRuleSpace
\if!\@collaboration!\else
{\Large\bfseries\sffamily\raggedright\@collaboration}\par
\afterCollaborationSpace
\fi
\if!\@collaborationImg!\else
{\normalsize\bfseries\sffamily\raggedright\@collaborationImg}\par
\afterCollaborationImgSpace
\fi
{\let\@footnotetext\H@@footnotetext
\bfseries\raggedright\sffamily\the\auth@toks\par}
\afterAuthorSpace
\ifaffil\begin{list}{}{%
\setlength{\leftmargin}{0.28cm}%
\setlength{\labelsep}{0pt}%
\setlength{\itemsep}{\affiliationsSep}%
\setlength{\topsep}{-\parskip}}
\itshape\small%
\the\affil@toks
\end{list}\fi
\afterAffiliationSpace
\ifemailadd\noindent\textbf{E-mail:} \the\email@toks\par\fi
\afterEmailSpace
\if!\@xtum!\else\noindent{\@xtum}\afterXtumSpace\fi
\if!\@abstract!\else\noindent{\renewcommand\baselinestretch{.9}\textsc{Abstract:}}\ \@abstract\afterAbstractSpace\fi
\if!\@keywords!\else\noindent{\textsc{Keywords:}} \@keywords\afterKeywordsSpace\fi
\if!\@arxivnumber!\else\noindent{\textsc{ArXiv ePrint:}} \href{https://arxiv.org/abs/\@arxivnumber}{\@arxivnumber}\afterArxivSpace\fi
\if!\@dedicated!\else\vbox{\small\it\raggedleft\@dedicated}\afterDedicatedSpace\fi
\ifnotoc\else
\iftoccontinuous\else\newpage\fi
\beforetochook\hrule
\tableofcontents
\afterTocSpace
\hrule
\afterTocRuleSpace
\fi
\setcounter{footnote}{0}
\pagestyle{myplain}\pagenumbering{arabic}
} %

\renewcommand{\baselinestretch}{1.1}\normalsize
\divide\@tempdima\baselineskip
\@tempcnta=\@tempdima
\skip\@mpfootins = \skip\footins
\renewcommand{\@dotsep}{10000}

\newcommand\ps@myplain{
\pagenumbering{arabic}
\renewcommand\@oddfoot{\hfill-- \thepage\ --\hfill}
\renewcommand\@oddhead{}}
\let\ps@plain=\ps@myplain

\newcommand\ps@titlepage{\renewcommand\@oddfoot{}\renewcommand\@oddhead{}}

\numberwithin{equation}{section}

\renewcommand\section{\@startsection{section}{1}{\z@}%
                                   {-3.5ex \@plus -1.3ex \@minus -.7ex}%
                                   {2.3ex \@plus.4ex \@minus .4ex}%
                                   {\normalfont\large\bfseries}}
\renewcommand\subsection{\@startsection{subsection}{2}{\z@}%
                                   {-2.3ex\@plus -1ex \@minus -.5ex}%
                                   {1.2ex \@plus .3ex \@minus .3ex}%
                                   {\normalfont\normalsize\bfseries}}
\renewcommand\subsubsection{\@startsection{subsubsection}{3}{\z@}%
                                   {-2.3ex\@plus -1ex \@minus -.5ex}%
                                   {1ex \@plus .2ex \@minus .2ex}%
                                   {\normalfont\normalsize\bfseries}}
\renewcommand\paragraph{\@startsection{paragraph}{4}{\z@}%
                                   {1.75ex \@plus1ex \@minus.2ex}%
                                   {-1em}%
                                   {\normalfont\normalsize\bfseries}}
\renewcommand\subparagraph{\@startsection{subparagraph}{5}{\z@}%
                                   {1.75ex \@plus1ex \@minus .2ex}%
                                   {-1em}%
                                   {\normalfont\normalsize\itshape}}

\def\fnum@figure{\textbf{\figurename\nobreakspace\thefigure}}
\def\fnum@table{\textbf{\tablename\nobreakspace\thetable}}

\long\def\@makecaption#1#2{%
  \vskip\abovecaptionskip
  \sbox\@tempboxa{\small #1. #2}%
  \ifdim \wd\@tempboxa >\hsize
    \small #1. #2\par
  \else
    \global \@minipagefalse
    \hb@xt@\hsize{\hfil\box\@tempboxa\hfil}%
  \fi
  \vskip\belowcaptionskip}

\renewenvironment{thebibliography}[1]{%
\begin{oldthebibliography}{#1}%
\small%
\raggedright%
\setlength{\itemsep}{5pt plus 0.2ex minus 0.05ex}%
}%
{%
\end{oldthebibliography}%
}

\makeatother
\usepackage[T1]{fontenc}
\usepackage[utf8]{inputenc}
\usepackage{amsmath,amssymb,mathtools}
\usepackage{booktabs}
\usepackage{array}
\usepackage{placeins}
\usepackage{tikz}
\usepackage{shuffle}
\usepackage{amsmath}
\renewcommand{\tablename}{Table}
\usepackage{needspace}
\usepackage{amsthm}
\newtheorem{lemma}{Lemma}
\newtheoremstyle{relationbody}
  {\topsep}{\topsep}{\itshape}{}{\bfseries}{}{0pt}{}
\theoremstyle{relationbody}
\newtheorem*{zrelation}{}
\newtheorem*{yrelation}{}
\newtheorem*{xrelation}{}
\theoremstyle{plain}
\usetikzlibrary{decorations.pathmorphing}
\newcommand{\YM}{\mathrm{YM}}
\newcommand{\NLSM}{\mathrm{NLSM}}
\newcommand{\cT}{\mathcal{T}}
\newcommand{\cL}{\mathcal{L}}
\newcommand{\cS}{\mathcal{S}}

\usetikzlibrary{
  arrows.meta,
  decorations.pathmorphing,
  decorations.markings
}

\tikzset{
  chfour figure/.style={
    font=\small,
    line width=.65pt,
    line cap=round,
    line join=round
  },
  chfour gluon/.style={
    draw,
    decorate,
    decoration={
      snake,
      amplitude=1.15pt,
      segment length=4.5pt
    }
  },
  chfour scalar/.style={
    draw,
    solid
  },
  chfour external/.style={
    draw,
    dash pattern=on 3pt off 2pt
  },
  chfour mixed/.style={
    draw,
    dash pattern=on 3pt off 2pt,
    postaction={decorate},
    decoration={
      markings,
      mark=at position .55 with {
        \arrow{Stealth[length=3.4pt,width=3pt]}
      }
    }
  },
  chfour vertex/.style={
    circle,
    fill=black,
    inner sep=1.3pt,
    outer sep=0pt
  },
  chfour momentum/.style={
    draw,
    solid,
    line width=.45pt,
    -{Stealth[length=3.5pt,width=3pt]}
  },
  chfour label/.style={
    inner sep=1.5pt,
    fill=white
  }
}
\title{One-loop unifying relations between Yang--Mills theory and the nonlinear sigma model from Berends--Giele currents}

\author[a]{Zhenhua Guo,}
\author[b]{Xin Yan}

\affiliation[a]{Arizona College of Technology, Hebei University of Technology,Tianjin 300401, China}
\affiliation[b]{Institute of Particle Physics and Key Laboratory of Quark and Lepton Physics (MOE), Central China Normal University, Wuhan 430079, China}

\emailAdd{zhenhuaguo1@arizona.edu}
\emailAdd{xinyan@mails.ccnu.edu.cn}

\abstract{We study unifying relations between Yang--Mills theory and the nonlinear sigma model. We extend tree-level amplitude relations expressed through differential operators to off-shell multiparticle currents and obtain one-loop integrand relations by sewing two off-shell legs. This off-shell approach may also apply to higher-loop integrands.
}


\begin{document}
\maketitle

\section{Introduction}
\label{sec:introduction}

Scattering amplitudes reveal structural connections between different quantum field theories that are not apparent at the level of their Lagrangians.
At tree level, amplitudes exhibit rich algebraic structures: color-ordered amplitudes satisfy the Kleiss--Kuijf (KK) and
Bern--Carrasco--Johansson (BCJ) relations~\cite{Kleiss:1988ne,Bern:2008qj},
while the Kawai--Lewellen--Tye (KLT) relations and color--kinematics duality reveal connections between gauge theories and gravity
~\cite{Kawai:1985xq,Bern:2008qj,Bern:2010ue}.
The scattering equations and the Cachazo--He--Yuan (CHY) representation provide a unified framework encompassing Yang--Mills (YM) theory, gravity, biadjoint scalar theory, and the nonlinear
$\sigma$ model (NLSM)
~\cite{Cachazo:2013hca,Cachazo:2013iea,Cachazo:2014xea}.
Amplitude unifying relations connect scattering amplitudes in different theories through differential operators~\cite{Cheung:2017ems,Zhou:2018wvn,Bollmann:2018edb} and constitute an important class of amplitude relations.

In recent years, progress has been made in several directions in extending tree-level amplitude relations to loop level.
One-loop KLT and BCJ relations have been studied~\cite{He:2016mzd,Stieberger:2022lss},
and color--kinematics duality and the double-copy structure have also been extended to one-loop and higher-loop amplitudes
~\cite{Boels:2013bi,Bern:2013yya,Bern:2019prr}.
Meanwhile, some tree-level unifying relations have been extended to one-loop Feynman integrands
~\cite{Zhou:2021gnu,Zhou:2022orv}.

Among the methods for extending tree-level constructions to higher loops, the off-shell approach is the most direct and concise: loop-level Feynman integrands are obtained by sewing off-shell tree-level objects~\cite{Gomez:2022dzk}\footnote{For the integral level and the celestial amplitude version, see \cite{Chattopadhyay:2021udc,Tao:2023wls,Chattopadhyay:2024kdq}.}. This off-shell sewing construction has also been extended to pure gravity, with a recursive treatment of diagram overcounting and symmetry factors~\cite{Gomez:2024GravityCorrelators}. However, computing and organizing off-shell objects is often cumbersome, calling for a description suited to recursion and sewing. We therefore use Berends--Giele multiparticle currents~\cite{Berends:1987me} to organize tree-level contributions into recursion relations derived from the classical equations of motion. These currents retain the tree-level information before the on-shell limit is taken and can be used to sew loop integrands. For a recent review of perturbiner methods and Berends--Giele currents, including NLSM recursions, see~\cite{Jusinskas:2026Perturbiner}. Existing work has mainly focused on on-shell methods, whereas studies that directly construct loop-level relations from off-shell tree-level relations~\cite{Chen:2013fya,Chen:2014dfa,Chen:2016off,Tao:2022AdSUnifying,Chen:2023dcx,Wu:2021exa,Tao:2023yxy,Tao:2024vcz} remain limited.

The relation between YM theory and the NLSM is an important example of amplitude unifying relations. Tree-level NLSM amplitudes can be obtained from higher-dimensional YM theory through generalized dimensional reduction~\cite{Cheung:2017yef}. Rewriting the nonlinear $\sigma$ model in a cubic representation involving three types of fields, $X$, $Y$, and $Z$, with only three-point interactions allows its off-shell tree-level objects to be constructed using multiparticle currents; this representation also makes flavor--kinematics duality manifest~\cite{Cheung:2016prv}. The tree-level unifying relation between the two theories has been verified using on-shell methods~\cite{Cheung:2017ems,Zhou:2018wvn,Bollmann:2018edb}. In this paper, we realize this tree-level unifying relation at the level of multiparticle currents and extend it to one-loop Feynman integrands by sewing, providing a concrete example of extending unifying relations to loop level through an off-shell approach.

The remainder of this paper is organized as follows. Section 2 introduces the basic definitions of transmutation operators and Berends--Giele multiparticle currents and presents the multiparticle-current recursion relations for YM theory and the cubic \(XYZ\) representation of the NLSM. Section 3 studies the transmutation relations between the two theories at the off-shell tree level, discussing the relation in section~\ref*{sec:y-current-relation} and the relation in section~\ref*{sec:x-current-relation} separately. Section 4 extends the tree-level current relations to one-loop integrands. We first show that diagrams containing four-point vertices give vanishing contributions after transmutation, and then test the one-loop unifying relation using bubble, tadpole, and triangle diagrams. Section 5 summarizes our main results and discusses possible extensions. The appendices provide proofs of the auxiliary identities used in the main text.

\section{Berends--Giele currents, transmutation operators, and off-shell sewing}
\label{sec:methods}

\subsection{Overview of unifying relations}
\label{sec:tree-operators}

The unifying relations considered here connect scattering amplitudes in different theories by differentiating with respect to Lorentz-invariant contractions. These differential operators are called transmutation operators. They preserve the KLT, BCJ, and CHY constructions of the theories, explaining why different theories share these constructions, and represent an important discovery in amplitude unifying relations.

Here, we treat $\epsilon_a\!\cdot\epsilon_b$, $\epsilon_a\!\cdot p_b$, and $p_a\!\cdot p_b$ as mutually independent formal variables. To maintain physical consistency, we require transmutation operators to preserve on-shell kinematics and gauge invariance. Three basic types of differential operators are considered: trace, longitudinal, and insertion operators. We do not discuss insertion operators here and use only the first two types.

Let $E,R$ be sets of labels, where $R$ specifies the polarization-vector labels on which the operators act and $E$ specifies the momentum labels included in the sum; $R\subseteq E$ is not required. We introduce the following notation~\cite{Cheung:2017ems,Zhou:2018wvn,Bollmann:2018edb}:
\begin{equation}
\begin{aligned}
 \cT_{ij}
 :=\frac{\partial}{\partial(\epsilon_i\!\cdot\epsilon_j)},
 \qquad
 &\cL_c^E
 :=\sum_{\substack{d\in E\\d\neq c}}
 (p_c\!\cdot p_d)
 \frac{\partial}{\partial(\epsilon_c\!\cdot p_d)},
 \qquad
 \cL_R^E:=\prod_{c\in R}\cL_c^E,\\
 &\cL\!\cdot\cT[i_1i_n]:=\cL_{i_2\cdots i_{n-1}}^{i_1i_2\cdots i_n}\cT_{i_1i_n}
 \label{eq:tree-operators}
 \end{aligned}
\end{equation}
The first operator is called the trace operator. It retains only terms containing $\epsilon_i\!\cdot\epsilon_j$ and removes this contraction. It can be realized by dimensional reduction, with the selected polarization vectors pointing along extra dimensions orthogonal to the gluon momenta, thereby transmuting a pair of gluons into a pair of scalars. The second operator is called the longitudinal operator. It identifies $\epsilon_c\!\cdot p_d$ term by term and replaces the polarization vector $\epsilon_c$ with the corresponding momentum $p_c$, thereby transmuting the particle into a longitudinal mode with derivative couplings. The definition of the longitudinal operator used here differs from the usual one: we omit the self term $p_c^2\,\partial/\partial(\epsilon_c\!\cdot p_c)$. Such terms, which vanish on shell, are allowed as redundancies in on-shell approaches, but they make the off-shell expressions less concise and are therefore omitted here.

Both trace and longitudinal operators contain partial derivatives with respect to Lorentz contractions involving at least one polarization vector: the trace operator contains $\frac{\partial}{\partial\epsilon_i\cdot \epsilon_j}$, and the longitudinal operator contains $\frac{\partial}{\partial p_i\cdot \epsilon_j}$. They act on functions of pure momentum-contraction variables as constants, so we focus only on the polynomial structure in contractions involving polarization vectors. The functions of formal variables encountered here, such as contractions of multiparticle currents and color-ordered partial amplitudes, are polynomials in formal variables involving polarization vectors, with coefficients in the field of rational functions of pure momentum contractions. Whenever we refer to polynomials and monomials, we mean polynomials and monomials in these polarization-dependent formal variables. For example, for the polynomial\[\epsilon_1\cdot \epsilon_2 +\frac{\epsilon_2\cdot p_3}{2\,s_{13}}+p_1\cdot p_2,\]
the monomials are $\epsilon_1\cdot \epsilon_2,\,\epsilon_2\cdot p_3,\,1$, with corresponding coefficients $1,\,\frac{1}{2\,s_{13}},\,p_1\cdot p_2$.

Since the trace and longitudinal operators commute with one another, $\cL_R^E$ depends only on the label sets $R$ and $E$, independently of the ordering of the labels within them.

In the unified web of theories described in ref.~\cite{Cheung:2017ems}, NLSM amplitudes can be obtained from YM amplitudes through transmutation operators:
\begin{equation}
 A_{\NLSM}=\cL\!\cdot\cT[i_1i_n]A_{\YM} .
 \label{eq:amplitude-unifying-relation}
\end{equation}
\subsection{YM multiparticle currents}
\label{sec:ym-bg}

We use the perturbiner method to derive the multiparticle currents of YM theory. Multiparticle currents depend on the gauge choice; we adopt Feynman gauge, $\xi=1$, and set the gauge coupling to $1$. The Feynman-gauge-fixed YM Lagrangian is~\cite{Faddeev:1967fc,Gomez:2022dzk,Lee:2022aiu,Chen:2023dcx}
\begin{equation}
\begin{aligned}
\mathcal{L}_{\mathrm{YM}}
&=
\operatorname{Tr}\!\left[
-\frac{1}{4}F_{\mu\nu}F^{\mu\nu}
-\frac{1}{2}(\partial_\mu A^\mu)^2
+\partial^\mu b\,D_\mu c
\right],
\\
F_{\mu\nu}
&=
\partial_\mu A_\nu
-\partial_\nu A_\mu
-[A_\mu,A_\nu],
\qquad
D_\mu c
=
\partial_\mu c-[A_\mu,c].
\end{aligned}
\label{eq:ym-lagrangian}
\end{equation}
Here, $b$ and $c$ are the Faddeev--Popov antighost and ghost fields, respectively. We mainly discuss pure-gluon tree-level amplitudes, so the multiparticle coefficients of the ghost fields are set to zero in proving the tree-level current relations. This simplification cannot be directly carried over to the one-loop correspondence, because the complete gauge-fixed YM one-loop integrand also includes contributions from closed ghost lines.

The Lagrangian~\eqref{eq:ym-lagrangian} gives the equations of motion
\begin{equation}
\begin{aligned}
\Box A^\mu
&=
-[A_\nu,F^{\mu\nu}]
+\partial_\nu[A^\nu,A^\mu]
-\{\partial^\mu b,c\},
\\
\Box b
&=
[A_\mu,\partial^\mu b],
\\
\Box c
&=
[A_\mu,\partial^\mu c]
+[\partial_\mu A^\mu,c].
\end{aligned}
\label{eq:ym-el}
\end{equation}

The perturbiner expansion takes the form~\cite{Selivanov:1999as,Lee:2015upy,Mafra:2015mja,Mizera:2018jbh}
\begin{equation}
\begin{aligned}
A^\mu(x)
&=
\sum_M A_M^\mu e^{p_M\cdot x}T^{a_M},
\\
b(x)
&=
\sum_M b_M e^{p_M\cdot x}T^{a_M},
\\
c(x)
&=
\sum_M c_M e^{p_M\cdot x}T^{a_M}.
\end{aligned}
\label{eq:ym-perturbiner}
\end{equation}
Here, $M=m_1m_2\cdots m_N$ is a sequence of letters, each letter $m_i$ representing an external leg. We call it an ordered word. Here, "ordered" does not mean that the subscripts are arranged in increasing order; rather, the order of the letters is part of the word itself. Different arrangements therefore define different words: for example, $m_1m_2$ and $m_2m_1$ are both ordered words, but they are not equal. We call $|M|=N$ the word length. For convenience, when no confusion can arise, we use $i$ to denote $m_i$; for example, $132$ denotes $m_1m_3m_2$. Note that the factor of $i$ has been absorbed into the momentum $p$. The total momentum of the multiparticle state is $p_M:=\sum_{r=1}^{N}p_{m_r}$, and we take all momenta to flow out of the vertex. The product $T^{a_M}:=T^{a_{m_1}}T^{a_{m_2}}\cdots T^{a_{m_N}}$ contains all color degrees of freedom of the multiparticle solution, where $a_{m_r}$ is the color index of external leg $m_r$, and $A_M^\mu$ is the color-stripped current coefficient. Whenever an ordered word appears in a set operation or a membership relation, it denotes the set of external-leg labels contained in that word; concatenation of words always preserves their order.

In this paper, a Berends--Giele (BG) current with more than one off-shell external leg is called a multiparticle current; the count of off-shell legs includes the root leg. The single-particle boundary condition is $A_a^{\mu}=\epsilon_a^\mu$, where $\epsilon_a^\mu$ is the polarization vector of external leg $a$. To construct multiparticle currents, we do not impose on-shell conditions on the single-particle inputs in advance, while the current coefficients with $|M|>1$ are required to satisfy the equations of motion~\eqref{eq:ym-el}.

Substituting eq.~\eqref{eq:ym-perturbiner} into the field equations obtained from eq.~\eqref{eq:ym-el} and matching the coefficients of $T^{a_M}$ gives the multiparticle-current recursion relations
\begingroup\small
\begin{equation}
\begin{aligned}
A_M^\mu
&=
\frac{1}{2s_M}
\Bigg\{
\sum_{M=QR}
\Bigg[
p_R^\mu b_R c_Q
+
\big((p_M+p_R)\cdot A_Q\big)A_R^\mu
+
\frac{1}{2}(p_Q-p_R)^\mu(A_Q\cdot A_R)
-
(Q\leftrightarrow R)
\Bigg]
\\
&\hspace{1.5cm}
+
\sum_{M=QRS}
\Bigg[
(A_Q\cdot A_S)A_R^\mu
-
(A_Q\cdot A_R)A_S^\mu
+
(Q\leftrightarrow S)
\Bigg]
\Bigg\}.
\end{aligned}
\label{eq:ym-bg-gluon-original}
\end{equation}
\endgroup

\begin{equation}
\begin{aligned}
b_M
&=
-\frac{1}{2s_M}
\sum_{M=QR}
\Big[
b_Q(p_Q\cdot A_R)
-
(Q\leftrightarrow R)
\Big],
\\
c_M
&=
-\frac{1}{2s_M}
\sum_{M=QR}
\Big[
c_Q(p_M\cdot A_R)
-
(Q\leftrightarrow R)
\Big].
\end{aligned}
\label{eq:ym-bg-ghost}
\end{equation}
Here, $s_M:=\frac12p_M^2$ is the Mandelstam invariant used in this paper. For an ordered word $M=m_1\cdots m_N$, the notation $M=QR$ denotes a split of $M$, in word order, into two nonempty contiguous subwords:
\[
Q=m_1\cdots m_r,\qquad R=m_{r+1}\cdots m_N,\qquad 1\le r<N.
\]
This is a binary deconcatenation: a split into two nonempty contiguous subwords. We refer to it below as a binary split. For example, for $M=1234$, the allowed splits are $(Q,R)=(1,234),(12,34),(123,4)$. A ternary deconcatenation $M=QRS$ likewise preserves the word order and requires all three contiguous subwords to be nonempty. All sums over splits in this paper follow this convention. The recursion can be interpreted in terms of tree diagrams: $1/2s_M$ is the propagator factor of the off-shell root leg; the three-gluon vertex terms are summed over binary splits, and the four-gluon vertex terms over ternary splits. The notation $(Q\leftrightarrow R)$ exchanges $Q$ and $R$ in all the preceding terms within the same square brackets; it originates from the commutators in the equations of motion.

At tree level, when all external legs are gluons, there are no internal ghost lines. Thus, in discussing tree-level unifying relations, we use the following recursion for pure-YM multiparticle currents:
\begin{equation}
\begin{aligned}
A_M^\mu
=
\frac{1}{2s_M}
\sum_{M=QR}
\Bigg[
&
2(A_Q\cdot p_R)A_R^\mu
+
(p_Q\cdot A_Q)A_R^\mu
-
(A_Q\cdot A_R)p_R^\mu
\\
&-
\sum_{R=TS}
\Big(
(A_Q\cdot A_T)A_S^\mu
-
(S\leftrightarrow T)
\Big)
-
(Q\leftrightarrow R)
\Bigg].
\end{aligned}
\label{eq:ym-bg-gluon}
\end{equation}

Interestingly, even when all external legs are off shell, one can prove that the multiparticle currents still satisfy the shuffle identity,\[
A^\mu_{Q\shuffle R}
\equiv
\sum_{P\in Q\shuffle R} A_P^\mu
=0.
\]where $Q\shuffle R$ denotes the sum of all shuffles of the words $Q$ and $R$ that preserve the internal order of each word, for example $12\shuffle 34=1234+1324+1342+3412+3124+3142$.

In YM theory, the BG formula for the color-ordered partial amplitude is
\begin{equation}
A_n(1,2,\ldots,n)
=
\lim_{p_n^2\to 0}
2s_{12\cdots n-1}\,
A_{12\cdots n-1}\cdot A_{n}.
\label{eq:ym-partial-amplitude}
\end{equation}

\subsection{The cubic \texorpdfstring{$X Y Z$}{XYZ} representation of the NLSM}
\label{sec:nlsm-bg}
The NLSM is an effective field theory describing the low-energy interactions of Goldstone bosons. Its color-ordered amplitudes satisfy KK, BCJ, and KLT-type relations
~\cite{Chen:2013fya,Cachazo:2014xea}. The conventional NLSM representation contains infinitely many interaction terms and is difficult to handle. Here, we use the cubic auxiliary-field representation of the NLSM, which is equivalent to the conventional representation at tree level~\cite{Cheung:2016prv,Cheung:2017yef,Mizera:2018jbh}:
\begin{equation}
 \mathcal L_{\NLSM}
 =\operatorname{Tr}\!\left[
 X_\mu\Box Z^\mu+\frac12Y\Box Y
 +\frac{1}{\sqrt2}
  (\partial_\mu X_\nu-\partial_\nu X_\mu)[Z^\mu,Z^\nu]
 +\frac{1}{\sqrt2}Z^\mu[Y,\partial_\mu Y]
 \right].
 \label{eq:xyz-lagrangian}
\end{equation}
This representation contains two adjoint vector fields, $X_\mu$ and $Z_\mu$, and one adjoint scalar field, $Y$, all Lie-algebra-valued and transforming in the adjoint representation. It leads to the following equations of motion:
\begin{equation}
\begin{aligned}
\Box X^\mu
&=
-\sqrt{2}\,[Z_\nu,\partial^\mu X^\nu-\partial^\nu X^\mu]
-\frac{1}{\sqrt{2}}\,[Y,\partial^\mu Y],
\\
\Box Y
&=
-\sqrt{2}\,[\partial_\mu Y,Z^\mu]
-\frac{1}{\sqrt{2}}\,[Y,\partial_\mu Z^\mu],
\\
\Box Z^\mu
&=
-\sqrt{2}\,\partial_\nu[Z^\mu,Z^\nu].
\end{aligned}
\label{eq:xyz-eom}
\end{equation}
The three fields share a common perturbiner expansion,
\begin{equation}
 \big(X^\mu,Y,Z^\mu\big)(x)
 =\sum_{P\neq\emptyset}
 \big(X_P^\mu,Y_P,Z_P^\mu\big)T^{a_P} e^{p_P\cdot x} .
 \label{eq:xyz-perturbiner}
\end{equation}
We first compute their fully off-shell multiparticle currents and then, for the configuration used in one-loop sewing, impose on-shell conditions on the external legs that are not sewn, in order to simplify the multiparticle-current recursion relations. Substituting eq.~\eqref{eq:xyz-perturbiner} into eq.~\eqref{eq:xyz-eom}, we consider only the current coefficients $X_M^{\mu}$, $Y_M$, and $Z_M^{\mu}$ with word length $|M|>1$, without imposing on-shell conditions on the single-particle states in advance. To facilitate establishing the unifying relation with YM theory, we replace the factor $1/\sqrt2$ at each three-point vertex with $1$.
The multiparticle-current recursion relations then become
\begin{equation}
\begin{aligned}
X_M^\mu
&=
-\frac{1}{s_M}
\sum_{M=QR}
\left[
p_R^\mu
\left(
\frac{1}{2}Y_QY_R+Z_Q\cdot X_R
\right)
+
(p_Q\cdot Z_R)X_Q^\mu
-
(Q\leftrightarrow R)
\right],\\
Y_M
&=
-\frac{1}{s_M}
\sum_{M=QR}
\left[
Y_Q
\left(
p_Q\cdot Z_R
+\frac{1}{2}p_R\cdot Z_R
\right)
-
(Q\leftrightarrow R)
\right],\\
Z_M^\mu
&=
-\frac{1}{s_M}
\sum_{M=QR}
\left[
Z_Q^\mu\,
(p_M\cdot Z_R)
-
(Q\leftrightarrow R)
\right]. 
\label{eq:recursion-offshell}
\end{aligned}
\end{equation}
Unlike the YM recursion, eq.~\eqref{eq:recursion-offshell} contains only binary splits, since the action~\eqref{eq:xyz-lagrangian} has only three-point interactions.
The formula for the color-ordered NLSM partial amplitude is now
\begin{equation}
A_n^{\mathrm{NLSM}}(1,2,\ldots,n)
=
\lim_{p_n^2\to 0}
2s_{12\cdots n-1}
\left(
X_{12\cdots n-1}^{\mu}Z_{n\mu}
+
Z_{12\cdots n-1}^{\mu}X_{n\mu}
+
Y_{12\cdots n-1}Y_n
\right).
\label{eq:nlsm-partial-amplitude}
\end{equation}
Previous studies have shown that only two types of external-state configurations yield nonzero amplitudes. Following their notation, we distinguish cases (a) and (b):
\begin{equation}
\begin{aligned}
\text{(a)}\qquad
&2\text{ }Y\text{-states and }n-2\text{ }Z\text{-states},
\\
\text{(b)}\qquad
&1\text{ } X\text{-state and }n-1\text{ }Z\text{-states}.
\end{aligned}
\label{eq:nlsm-nontrivial-states}
\end{equation} 
The corresponding single-particle external states are
\begin{subequations}
\label{eq:nlsm-boundary-conditions}
\begin{align}
\text{(a)}\qquad
(X_i^\mu,Y_i,Z_i^\mu)
&=
\begin{cases}
(0,1,0), & i\text{ is a }Y\text{-state},\\
(0,0,p_i^\mu), & i\text{ is a }Z\text{-state},
\end{cases}
\label{eq:nlsm-boundary-YZ}
\\
\text{(b)}\qquad
(X_i^\mu,Y_i,Z_i^\mu)
&=
\begin{cases}
(\bar{p}_i^\mu,0,0), & i\text{ is an }X\text{-state},\\
(0,0,p_i^\mu), & i\text{ is a }Z\text{-state}.
\end{cases}
\label{eq:nlsm-boundary-XZ}
\end{align}
\end{subequations}
Here, $\bar{p}_i^\mu$ is an auxiliary vector, conventionally required to satisfy $\bar{p}_i\cdot p_i=-1$. Since we are discussing multiparticle currents, however, we regard $\bar{p}_i^{\mu}$ as an unconstrained independent variable. We use $\overline{i}$ to denote that leg \(i\) is an \(X\)-state, $\underline{i}$ that leg \(i\) is a \(Y\)-state, and $i$ that leg \(i\) is a \(Z\)-state. In cases (a) and (b), the values do not depend on which legs are chosen as the $Y$-states or the $X$-state, reflecting Bose symmetry.

The correspondence of the $X$ and $Z$ external states with NLSM amplitudes is established when the external legs satisfy the on-shell conditions $p_i^2=0$. Since one-loop sewing turns external legs into off-shell internal lines, the original external-state correspondence cannot be directly extended to the internal states after sewing. Thus, although the $X$--$Z$ sewing contributions belong to the one-loop integrand of the complete auxiliary-field theory, they do not correspond to the NLSM one-loop integrand. By contrast, the off-shell current relations in the $Y$ sector remain valid during sewing, so only $Y$--$Y$ sewing produces the integrand corresponding to NLSM one-loop amplitudes~\cite{Edison:2023xyz}. Accordingly, in this paper we consider only the tree-level unifying relations for external-state configuration (a) and the loop-level unifying relations obtained by sewing its two $Y$ legs.

Since we sew only the two $Y$ legs, we consider multiparticle currents with two off-shell $Y$ external legs and all remaining $Z$ legs on shell. For these multiparticle currents, an on-shell $Z$ external state obeys $Z_a\cdot p_a=0$, while the two $Y$ legs themselves have $Z_i^\mu=0$. We also observe that, for $|M|>1$, $Z_M\cdot p_M=0$ holds without imposing on-shell conditions on the individual legs. This follows immediately by expanding it using eq.~\eqref{eq:recursion-offshell}:
\[Z_M\cdot p_M
 =-\frac{1}{s_M}\sum_{M=QR}
 \Big[
 (Z_Q\cdot p_M)( Z_R\!\cdot p_M)-(Q\leftrightarrow R)
 \Big]
 =0.\]

The multiparticle-current recursion relations can therefore be simplified to
\begin{subequations}
\label{eq:recursion}
\begin{align}
X_M^\mu
&=
-\frac{1}{s_M}
\sum_{M=QR}
\left[
p_R^\mu
\left(
\frac{1}{2}Y_QY_R+Z_Q\cdot X_R
\right)
+
(p_Q\cdot Z_R)X_Q^\mu
-
(Q\leftrightarrow R)
\right],
\label{eq:recursion-X}
\\
Y_M
&=
-\frac{1}{s_M}
\sum_{M=QR}
\left[
Y_Q(p_Q\cdot Z_R)
-
(Q\leftrightarrow R)
\right],
\label{eq:recursion-Y}
\\
Z_M^\mu
&=
-\frac{1}{s_M}
\sum_{M=QR}
\left[
Z_Q^\mu(p_Q\cdot Z_R)
-
(Q\leftrightarrow R)
\right].
\label{eq:recursion-Z}
\end{align}
\end{subequations}
Under the external-leg conditions above, these have the same form as the usual NLSM BG recursion.

\section{Tree-level off-shell relations between YM and NLSM currents}
\label{sec:tree-relations}

For external-state configuration (a), we take all $Y$ legs in $M$ off shell and all $Z$ legs in $M$ on shell, while the additional leg $n\notin M$ is off shell whether it is a $Y$ or a $Z$ state. On the YM side, the corresponding gluon legs in $M$ have the same on- or off-shell assignments. We consider two choices of legs for the trace operator. If it selects $i\in M$ and the additional leg $n$, applying the transmutation operator yields $Y_MY_n$ on the NLSM side, giving the $Y$-current relation; if it selects $i,j\in M$, the NLSM side contains $X_M\cdot Z_n$ together with an off-shell correction involving $\Gamma_M$, giving the $X$-current relation. Taking the $Z$ external legs in $M$ on shell allows us to use eq.~\eqref{eq:recursion} as the multiparticle-current recursion. Apart from this use, no additional on-shell conditions or momentum conservation are required.

More generally, for this configuration we find the following correspondences between the multiparticle currents of YM theory and the NLSM:
\begin{enumerate}
    \item If $M$ contains $0$ $Y$-states and all remaining states are $Z$-states, then $A_M\rightarrow Z_M,$ as guaranteed by the relation in section~\ref*{sec:z-current-relation}.
    \item If $M$ contains $1$ $Y$-state and all remaining states are $Z$-states, then $A_M\rightarrow Y_M,$ as guaranteed by the relation in section~\ref*{sec:y-current-relation}.
    \item If $M$ contains $2$ $Y$-states and all remaining states are $Z$-states, then $A_M\rightarrow X_M,$ as guaranteed by the relation in section~\ref*{sec:x-current-relation}.
\end{enumerate}
We first introduce some basic properties that apply generally to off-shell unifying relations and prepare for the rigorous derivations below.

Polarization vectors play a crucial role in YM theory, and amplitudes are multilinear in them. This property is physically natural because a scattering amplitude is linear in each external state. For rigor, it can be proved as follows. Multilinearity means that each polarization vector appears exactly once through contraction variables in every monomial of the amplitude. This follows by analyzing the YM multiparticle-current recursion: $A_M\cdot A_n$ is manifestly linear in $A_n=\epsilon_n$. For $A_M$, the sum over lower-point currents on the right-hand side of eq.~\eqref{eq:ym-bg-gluon} shows that if the lower-point currents are linear in all their polarization vectors, then so is $A_M$. The single-particle boundary condition $A_i=\epsilon_i$ is manifestly linear in $\epsilon_i$. Mathematical induction therefore establishes that $A_M$ is linear in every $\epsilon_i$ with $i\in M$.

We can also establish the following property of YM multiparticle currents: if a monomial in \(A_M^\mu\), after removing the external-leg variable carrying the index $\mu$, contains no contraction of two polarization vectors, then the removed variable carrying the index $\mu$ is $\epsilon_k^\mu$ for some $k\in M$. This can be proved by induction. We briefly outline the proof. The recursion~\eqref{eq:ym-bg-gluon} shows that the expansion of \(A_P^\mu\) contains
\[
(A_R\cdot p_S)A_S^\mu,\qquad
p_S^\mu(A_R\cdot A_S),\qquad
(A_Q\cdot A_R)A_S^\mu,\qquad
(A_R\cdot A_S)A_Q^\mu
\]four types of terms. If the property under consideration holds for a polynomial, it continues to hold when that polynomial is multiplied by any other polynomial. Thus, it suffices to prove it for these four types. For the first type, the property follows from the lower-point current $A_S^\mu$. For the second type, removing the external-leg variable $p_S$ carrying the index $\mu$ leaves the polynomial $(A_R\cdot A_S)$; if the lower-point currents satisfy the property, their contraction naturally does as well. The third and fourth types are treated in the same way as the second. Thus, all four types inherit the property from lower-point currents, so it suffices to verify the lowest-point case. The single-particle current manifestly has this property.

The property above implies that every monomial in the contraction of two multiparticle currents contains a contraction of two polarization vectors. Consider two multiparticle currents $A_M$ and $A_N$ with disjoint words, $M\cap N=\varnothing$. By multilinearity in the polarization vectors, each polarization vector can appear only once in a monomial. The disjointness of $M$ and $N$ also means that $A_M^{\mu}$ and $A_N^{\mu}$ contain distinct sets of polarization vectors. Hence, if a monomial in $A_M\cdot A_N$ contains $\epsilon_i\cdot \epsilon_j$, it cannot also contain the momentum--polarization contractions $p_k\cdot \epsilon_i$ or $p_k\cdot \epsilon_j$. The longitudinal operators $\cL_i$ and $\cL_j$ with the corresponding labels therefore annihilate that monomial. It follows that

\begin{equation}
\cL^{M\cup N}_{M\cup N}
(A_M\cdot A_N)=0.
\label{eq:LM-AMAN-zero}
\end{equation}
This property ensures that Feynman diagrams containing four-point vertices vanish under the transmutation operator. The tree-diagram topologies of the two theories are therefore identical.

When a product of these operators acts on a formal polynomial multilinear in the polarization vectors, multilinearity immediately implies that the result vanishes if a trace operator and a longitudinal operator involve the same polarization vector. The result also vanishes if the operator product involves more distinct polarization vectors than are present in the formal polynomial. These properties continue to hold when the operators act on a product of formal polynomials, since such a product is itself a formal polynomial. If the different factors contain disjoint sets of polarization vectors, the operators in the product can be moved, according to their polarization labels, to act on the corresponding factors. We call this the distribution rule for operator products. For example, $\cL_1\cL_2\cT_{45}\cL_3A_{12}\cdot A_3 A_{4}\cdot A_5=(\cL_1\cL_2\cL_3A_{12}\cdot A_3)(\cT_{45}A_{4}\cdot A_5)$.

The difference between the multiparticle-current recursions in Feynman gauge and Lorenz gauge lies in the presence of terms such as $A_M\cdot p_M$. We can prove that such terms vanish under transmutation operators. Let $M$ and $N$ be nonempty and disjoint.
\begin{equation}
    \begin{aligned}
        \cL_{M}^{M\cup N}A_M\cdot p_M=0.
    \end{aligned}
    \label{eq:ampm}
\end{equation}
We proceed by induction on the word length $|M|$. When $|M|=1$, let $M=\{j\}$. Using the single-particle boundary condition $A_j^\mu=\epsilon_j^\mu$, we obtain
\[
\begin{aligned}
\cL_j^{j\cup N}(A_j\cdot p_j)
=\sum_{a\in N}p_j\cdot p_a\,\partial_{p_a\cdot \epsilon_j}(\epsilon_j\cdot p_j)=0,
\end{aligned}
\]
so the statement holds for $|M|=1$. Assume that, for every word $K$ of length less than $|M|$,
\[ \cL_{K}^{K\cup N}A_K\cdot p_K=0.\]

For a word $M$ of length $|M|>1$, the YM multiparticle-current recursion~\eqref{eq:ym-bg-gluon} gives
\begin{equation}
\begin{aligned}
A_M\cdot p_M
=
\frac{1}{2s_M}
\sum_{M=QR}
\Bigg[
&\underbrace{
2(A_Q\cdot p_R)(A_R\cdot p_M)
}_{\mathrm{(I)}}
+
\underbrace{
(p_Q\cdot A_Q)(A_R\cdot p_M)
}_{\mathrm{(II)}}
-
\underbrace{
(A_Q\cdot A_R)(p_R\cdot p_M)
}_{\mathrm{(III)}}
\\
&-
\sum_{R=TS}
\Big(
\underbrace{(A_Q\cdot A_T)(A_S\cdot p_M)}_{\mathrm{(IV)}}
-(S\leftrightarrow T)
\Big)-(Q\leftrightarrow R)
\Bigg].
\end{aligned}
\end{equation}
Apply $\cL_M^{M\cup N}$ to both sides. We analyze the contributions of the four types of terms, $\mathrm{(I)}$--$\mathrm{(IV)}$, separately. For $\mathrm{(I)}$, since $Q\cap R=\varnothing$, the polarization variables in the first factor come only from $Q$, and those in the second only from $R$. Thus,
\begingroup\small
\begin{align*}
    \cL_M^{M\cup N}\!\left[(A_Q\cdot p_R)(A_R\cdot p_M)\right]&=\cL_{Q}^{M\cup N}(A_Q\cdot p_R)\,\cL_R^{M\cup N}(A_R\cdot p_M)\\
    =&\cL_{Q}^{M\cup N}(A_Q\cdot p_R)\,\cL_R^{M\cup N}(A_R\cdot p_R)+\cL_{Q}^{M\cup N}(A_Q\cdot p_R)\,\cL_R^{M\cup N}(A_R\cdot p_Q)\\
    =&\cL_{Q}^{M\cup N}(A_Q\cdot p_R)\,\cL_R^{M\cup N}(A_R\cdot p_Q).
\end{align*}
\endgroup
Here, we used the induction hypothesis $\cL_R^{M\cup N}(A_R\cdot p_R)=0$. Since $A_R\cdot p_R$ contains neither the polarization vectors nor the momenta associated with $Q$, the operators $\cL_R^{M\cup N}$ and $\cL_R^{R\cup N}$ have the same action on it. We will use this replacement directly in analogous cases below without repeating the explanation. Including the $-(Q\leftrightarrow R)$ term associated with $\mathrm{(I)}$ gives an exact cancellation, so the contribution vanishes. The distribution rule for operator products is used frequently in these proofs. We will therefore not explain it repeatedly: when a reminder is useful, we will state that the distribution rule has been used, and when its use is evident, we will leave it implicit.

For $\mathrm{(II)}$, the distribution rule for operator products gives
\[\cL_M^{M\cup N}\!\left[(p_Q\cdot A_Q)(A_R\cdot p_M)\right]=\cL_{Q}^{Q\cup N}(p_Q\cdot A_Q)\,\cL_R^{R\cup M}(A_R\cdot p_M)=0.\]
For $\mathrm{(III)}$, using the distribution rule for operator products gives
\[
\cL_M^{M\cup N}\!\left[(A_Q\cdot A_R)(p_R\cdot p_M)\right]
=\left(\cL_M^{M\cup N}(A_Q\cdot A_R)\right)(p_R\cdot p_M)=0.
\]
The last step uses eq.~\eqref{eq:LM-AMAN-zero}. This readily recognizable property is used frequently in the proofs below; for brevity, we do not cite it every time. Thus, $\mathrm{(III)}$ gives no contribution.

For $\mathrm{(IV)}$, the distribution rule for operator products gives
\[
\begin{aligned}
\cL_M^{M\cup N}\!\left[(A_Q\cdot A_T)(A_S\cdot p_M)\right]
=\cL_{QT}^{QT}(A_Q\cdot A_T)\,\cL_S^{S\cup N}(A_S\cdot p_M)=0.
\end{aligned}
\]
The corresponding $Q\leftrightarrow R$ or $S\leftrightarrow T$ terms of $\mathrm{(II)}$--$\mathrm{(IV)}$ have exactly the same structure and therefore also vanish. This completes the inductive proof of the property.
\subsection{Lemma: \texorpdfstring{$Z$}{Z}-current relation}
\label{sec:z-current-relation}
We now state the $Z$-current relation.
\begin{zrelation}
\phantomsection\label{lem:z-current}
Let $N$ be a label set disjoint from $M$, with all labels in $M$ corresponding to $Z$-states and $N$ nonempty. We observe that
\begin{equation}
 \cL_M^{M\cup N}\big(A_M\!\cdot p_N\big)
 =Z_M\!\cdot p_N .
 \label{eq:z-current-lemma}
\end{equation}
\end{zrelation}
We proceed by induction on the word length $|M|$. When $|M|=1$, let $M=\{j\}$, where $j$ is a $Z$-state.
Then
\[
\begin{aligned}
\cL_j^{j\cup N}(A_j\cdot p_N)
&=\cL_j^{j\cup N}(\epsilon_j\cdot p_N)
=\sum_{a\in N}p_j\cdot p_a\,\partial_{p_a\cdot \epsilon_j}(\epsilon_j\cdot p_N)\\
&=p_j\cdot p_N=Z_j\cdot p_N.
\end{aligned}
\]
Thus, the statement holds for $|M|=1$. Assume that, for every word $K$ of length less than $|M|$ and every label set $N'$ disjoint from $K$,
\[
\cL_K^{K\cup N'}(A_K\cdot p_{N'})=Z_K\cdot p_{N'}.
\]
For a word $M$ of length $|M|>1$, the YM multiparticle-current recursion~\eqref{eq:ym-bg-gluon}
gives
\begin{equation}
\begin{aligned}
A_M\cdot p_N
=
\frac{1}{2s_M}
\sum_{M=QR}
\Bigg[
&\underbrace{
2(A_Q\cdot p_R)(A_R\cdot p_N)
}_{\mathrm{(I)}}
+
\underbrace{
(p_Q\cdot A_Q)(A_R\cdot p_N)
}_{\mathrm{(II)}}
-
\underbrace{
(A_Q\cdot A_R)(p_R\cdot p_N)
}_{\mathrm{(III)}}
\\
&-
\sum_{R=TS}
\Big(
\underbrace{(A_Q\cdot A_T)(A_S\cdot p_N)}_{\mathrm{(IV)}}
-(S\leftrightarrow T)
\Big)-(Q\leftrightarrow R)
\Bigg].
\end{aligned}
\label{eq:ym-bg-contracted}
\end{equation}
Apply $\cL_M^{M\cup N}$ to both sides of eq.~\eqref{eq:ym-bg-contracted}. We analyze the terms separately:

$\mathrm{(I)}$:
\[
\qquad \cL_M^{M\cup N}\!\left[(A_Q\cdot p_R)(A_R\cdot p_N)\right]
=\cL_Q^{Q\cup R}(A_Q\cdot p_R)\,\cL_R^{R\cup N}(A_R\cdot p_N).
\]

$\mathrm{(II)}$:\[ \cL_M^{M\cup N}\!\left[(p_Q\cdot A_Q)(A_R\cdot p_N)\right]=\cL_{Q}^{Q}(p_Q\cdot A_Q)\,\cL_R^{R\cup N}(A_R\cdot p_N)=0.\]
The second equality uses eq.~\eqref{eq:ampm}.

$\mathrm{(III)}$:\[ 
\cL_M^{M\cup N}\!\left[(A_Q\cdot A_R)(p_R\cdot p_N)\right]
=\left(\cL_M^{M\cup N}(A_Q\cdot A_R)\right)(p_R\cdot p_N)=0.
\]

$\mathrm{(IV)}$:
\[
\cL_M^{M\cup N}\!\left[(A_Q\cdot A_T)(A_S\cdot p_N)\right]
=\cL_{QT}^{QT}(A_Q\cdot A_T)\,\cL_S^{S\cup N}(A_S\cdot p_N)=0.
\]
The corresponding $Q\leftrightarrow R$ or $S\leftrightarrow T$ terms likewise vanish. Thus, only $\mathrm{(I)}$ and its $Q\leftrightarrow R$ term contribute.

Since $M=QR$ is a binary split,
$
0<|Q|<|M|,\, 0<|R|<|M|,
$
and since $M$ consists entirely of $Z$-states and $M\cap N=\varnothing$, both $Q$ and $R$ consist entirely of $Z$-states, with $Q\cap R=Q\cap N=R\cap N=\varnothing.$
The induction hypothesis can therefore be applied to each factor:
\[
\cL_Q^{Q\cup R}(A_Q\cdot p_R)=Z_Q\cdot p_R,\qquad
\cL_R^{R\cup N}(A_R\cdot p_N)=Z_R\cdot p_N.
\]
For the exchanged term, we similarly have
\[
\cL_R^{R\cup Q}(A_R\cdot p_Q)=Z_R\cdot p_Q,\qquad
\cL_Q^{Q\cup N}(A_Q\cdot p_N)=Z_Q\cdot p_N.
\]
Substituting back into eq.~\eqref{eq:ym-bg-contracted} gives
\[
\begin{aligned}
\cL_M^{M\cup N}(A_M\cdot p_N)
&=\frac{1}{s_M}\sum_{M=QR}
\Big[(Z_Q\cdot p_R)(Z_R\cdot p_N)-(Z_R\cdot p_Q)(Z_Q\cdot p_N)\Big]\\
&=\left[-\frac{1}{s_M}\sum_{M=QR}
\Big((Z_R\cdot p_Q)Z_Q^\mu-(Z_Q\cdot p_R)Z_R^\mu\Big)\right]p_{N\mu}\\
&=Z_M\cdot p_N,
\end{aligned}\]
The last step uses the multiparticle recursion~\eqref{eq:recursion-Z} for the NLSM $Z$-current. This completes the inductive proof of the relation in section~\ref*{sec:z-current-relation}.

\subsection{\texorpdfstring{$Y$}{Y}-current relation}
\label{sec:y-current-relation}
We now consider the tree-level unifying relation between YM theory and the NLSM for external-state configuration $(a)$.
\begin{yrelation}
\phantomsection\label{lem:y-current}
Let external legs \(n\notin M\) and \(i\in M\) be \(Y\)-states, and let all external legs in \(M\setminus i\) be \(Z\)-states. We observe that
\begin{equation}
\cL_{M\setminus i}\cT_{in}\big(A_M\cdot A_n\big)=Y_M Y_n .
\label{eq:y-current-lemma}
\end{equation}
\end{yrelation}

We proceed by induction on the word length \(|M|\). When \(|M|=1\), \(M=\{i\}\), and
\[
\cL_{M\setminus i}\cT_{in}(A_i\cdot A_n)
=\cT_{in}(\epsilon_i\cdot \epsilon_n)=1=Y_iY_n.
\]
Thus, the relation in section~\ref*{sec:y-current-relation} holds for \(|M|=1\). Assume that the relation in section~\ref*{sec:y-current-relation} holds for every word \(K\) of length less than \(|M|\). For \(|M|>1\), since \(A_n^\mu=\epsilon_n^\mu\), the YM multiparticle-current recursion~\eqref{eq:ym-bg-gluon} gives
\begin{equation}
\begin{aligned}
A_M\cdot \epsilon_n
=
\frac{1}{2s_M}
\sum_{M=QR}
\Bigg[
&\underbrace{
2(A_Q\cdot p_R)(A_R\cdot \epsilon_n)
}_{\mathrm{(I)}}
+
\underbrace{
(p_Q\cdot A_Q)(A_R\cdot \epsilon_n)
}_{\mathrm{(II)}}
-
\underbrace{
(A_Q\cdot A_R)(p_R\cdot \epsilon_n)
}_{\mathrm{(III)}}
\\
&-
\sum_{R=TS}
\Big(
\underbrace{
(A_Q\cdot A_T)(A_S\cdot \epsilon_n)
}_{\mathrm{(IV)}}
-(S\leftrightarrow T)
\Big)
-(Q\leftrightarrow R)
\Bigg].
\end{aligned}
\label{eq:ym-bg-contracted-en}
\end{equation}
Apply \(\cL_{M\setminus i}\cT_{in}\) to both sides of eq.~\eqref{eq:ym-bg-contracted-en}. We analyze \(\mathrm{(I)}\)--\(\mathrm{(IV)}\) separately.

For \(\mathrm{(II)}\), \(\cT_{in}\) can give a nonzero result when acting on \(A_R\cdot \epsilon_n\) only if \(i\in R\). In this case, \(i\notin Q\), and the operator product is distributed as
\[
\begin{aligned}
\cL_{M\setminus i}\cT_{in}
\left[(p_Q\cdot A_Q)(A_R\cdot \epsilon_n)\right]
&=
\cL_Q^{M\setminus i}(p_Q\cdot A_Q)\,
\cL_{R\setminus i}\cT_{in}(A_R\cdot \epsilon_n)
=0,
\end{aligned}
\]
The last step uses eq.~\eqref{eq:ampm}. If \(i\notin R\), then \(A_R\cdot \epsilon_n\) contains no \(\epsilon_i\), so \(\cT_{in}\) directly gives zero. Thus, \(\mathrm{(II)}\) gives no contribution.

For \(\mathrm{(III)}\), \(A_Q\cdot A_R\) contains no \(\epsilon_n\), while \(p_R\cdot \epsilon_n\) contains no \(\epsilon_i\). Hence,
\[
\cT_{in}\left[(A_Q\cdot A_R)(p_R\cdot \epsilon_n)\right]=0,
\]
so \(\mathrm{(III)}\) gives no contribution.

For \(\mathrm{(IV)}\), if \(i\notin S\), then \(A_S\cdot \epsilon_n\) contains no \(\epsilon_i\), and \(\cT_{in}\) directly gives zero. If \(i\in S\), then \(i\notin Q\cup T\), and
\[
\begin{aligned}
\cL_{M\setminus i}\cT_{in}
\left[(A_Q\cdot A_T)(A_S\cdot \epsilon_n)\right]
&=
\cL_{QT}^{QT}(A_Q\cdot A_T)\,
\cL_{S\setminus i}\cT_{in}(A_S\cdot \epsilon_n)
=0,
\end{aligned}
\]
The last step uses eq.~\eqref{eq:LM-AMAN-zero}. Its \(S\leftrightarrow T\) term is treated identically. Thus, \(\mathrm{(IV)}\) also gives no contribution. The corresponding \(Q\leftrightarrow R\) terms of \(\mathrm{(II)}\)--\(\mathrm{(IV)}\) vanish for exactly the same reasons, so only \(\mathrm{(I)}\) and its \(Q\leftrightarrow R\) term can give nonzero contributions.

Now consider \(\mathrm{(I)}\). If \(i\in R\), then \(i\notin Q\), so \(Q\) is a pure-\(Z\) subword, and
\begin{equation}
\begin{aligned}
\cL_{M\setminus i}\cT_{in}
\left[(A_Q\cdot p_R)(A_R\cdot \epsilon_n)\right]
&=
\cL_Q^{Q\cup R}(A_Q\cdot p_R)\,
\cL_{R\setminus i}\cT_{in}(A_R\cdot \epsilon_n).
\end{aligned}
\label{eq:y-current-I-R}
\end{equation}
Since \(Q\cap R=\varnothing\) and all external legs in \(Q\) are \(Z\)-states, the relation in section~\ref*{sec:z-current-relation} gives
\[
\cL_Q^{Q\cup R}(A_Q\cdot p_R)=Z_Q\cdot p_R.
\]
Since \(|R|<|M|\), the induction hypothesis gives
\[
\cL_{R\setminus i}\cT_{in}(A_R\cdot \epsilon_n)=Y_RY_n.
\]
Therefore, when \(i\in R\), the contribution of \(\mathrm{(I)}\) is
\[
(Z_Q\cdot p_R)Y_RY_n.
\]

On the other hand, if \(i\in Q\), the original \(\mathrm{(I)}\) term vanishes, while its \(Q\leftrightarrow R\) term can contribute. In this case, \(i\notin R\), so \(R\) is a pure-\(Z\) subword, and
\[
\begin{aligned}
\cL_{M\setminus i}\cT_{in}
\left[(A_R\cdot p_Q)(A_Q\cdot \epsilon_n)\right]
&=
\cL_R^{R\cup Q}(A_R\cdot p_Q)\,
\cL_{Q\setminus i}\cT_{in}(A_Q\cdot \epsilon_n)
\\
&=(Z_R\cdot p_Q)Y_QY_n,
\end{aligned}
\]
Here, we have used the relation in section~\ref*{sec:z-current-relation} and the induction hypothesis, respectively. This term comes from \(-(Q\leftrightarrow R)\) in eq.~\eqref{eq:ym-bg-contracted-en} and therefore carries a minus sign.

Substituting \(\mathrm{(I)}\) and its exchanged term back into eq.~\eqref{eq:ym-bg-contracted-en}, and using the factor of \(2\) in front of them to cancel the overall \(1/2\), gives
\begin{equation}
\begin{aligned}
\cL_{M\setminus i}\cT_{in}(A_M\cdot \epsilon_n)
&=
\frac{Y_n}{s_M}
\sum_{M=QR}
\Big[
(Z_Q\cdot p_R)Y_R
-(Z_R\cdot p_Q)Y_Q
\Big]
\\
&=
-\frac{Y_n}{s_M}
\sum_{M=QR}
\Big[
Y_Q(Z_R\cdot p_Q)
-(Q\leftrightarrow R)
\Big]
\\
&=Y_MY_n .
\end{aligned}
\label{eq:y-current-induction}
\end{equation}
The last step uses the recursion for the NLSM \(Y\)-current, eq.~\eqref{eq:recursion-Y}. If \(Q\) or \(R\) does not contain the unique \(Y\) external leg \(i\), that subword is a pure \(Z\) subword and its corresponding \(Y_Q\) or \(Y_R\) automatically vanishes (as established by lemma~\ref{lem:even-non-z-y-zero}). Thus, eq.~\eqref{eq:y-current-induction} already includes both cases, \(i\in Q\) and \(i\in R\). This completes the induction and proves the relation in section~\ref*{sec:y-current-relation}.

We now prove that, if all external legs in the word \(M\) are \(Z\)-states, then
$
Y_M=0.
\label{eq:pure-z-y-zero}
$
We state the following two lemmas:

\begin{lemma}
\label{lem:non-z-z-zero}
If the word $M$ contains a nonzero number of non-$Z$ states, then $Z_M^\mu=0.$
\end{lemma}

When $|M|=1$, the only possibilities are $M={\underline{i}}$ or ${\overline{i}}$, both of which have $Z_M=0$. Assume that the statement holds for words of length less than $|M|$. The recursion is
\[Z_M^\mu = -\frac{1}{s_M} \sum_{M=QR} \left(Z_Q^\mu p_Q\cdot Z_R - (Q \leftrightarrow R) \right) \] 
Here, $|Q|,|R|<|M|$. Since $Q,R$ are the two nonempty contiguous subwords in the binary split $M=QR$, at least one of them contains a non-$Z$ state. The induction hypothesis gives $Z_Q=0$ or $Z_R=0$. Substitution into the recursion immediately yields $Z_M^\mu=0.$ This proves the lemma.

\begin{lemma}
\label{lem:even-non-z-y-zero}
If the number of non-\(Z\) external states in the word $M$ is even, then $Y_M=0.$
\end{lemma}
           
When $|M|=1$, the only possibility for $M$ is $\{a\}$ with $a$ a $Z$-state, and $Y_a=0$.
Assume that the statement holds for words of length less than $|M|$. The recursion is
 \[Y_M
=
-\frac{1}{s_M}
\sum_{M=QR}
\left(
Y_Q p_Q\cdot Z_R
-
Y_R p_R\cdot Z_Q \right)\]
The words $Q,R$ are the two nonempty contiguous subwords in the binary split $M=QR$. Since the total number of non-$Z$ external states in $M$ is even, the numbers of non-$Z$ external states in $Q,R$ have the same parity: both are even or both are odd. In the former case, the induction hypothesis gives $Y_Q=Y_R=0$; in the latter, lemma~\ref{lem:non-z-z-zero} gives $Z_Q^\mu=Z_R^\mu=0$. Both cases imply $Y_M=0,$ proving the lemma. In particular, if $M$ is a pure-$Z$ word, lemma~\ref{lem:even-non-z-y-zero} gives $Y_M=0.$

We give a simple example to verify the statement above. Take $M=123$, $i=2$, and $n=4$. The YM multiparticle-current recursion~\eqref{eq:ym-bg-contracted-en} gives
\begingroup\small
\begin{align}
A_{123}\cdot A_4
&=
\frac{1}{2s_{123}}
\Bigg[
\bigg[
\underbrace{
2(\epsilon_1\cdot p_{23})(A_{23}\cdot \epsilon_4)
}_{\mathrm{(I)}\,2(A_1\cdot p_{23})(A_{23}\cdot A_4)}
+
\underbrace{
(p_1\cdot \epsilon_1)(A_{23}\cdot \epsilon_4)
}_{\mathrm{(II)}\,(p_1\cdot A_1)(A_{23}\cdot A_4)}
-
\underbrace{
(\epsilon_1\cdot A_{23})(p_{23}\cdot \epsilon_4)
}_{\mathrm{(III)}\,(A_1\cdot A_{23})(p_{23}\cdot A_4)}
-(Q\leftrightarrow R)
\bigg]
\notag\\
&\quad
-
\underbrace{
(\epsilon_1\cdot \epsilon_2)(\epsilon_3\cdot \epsilon_4)
-(\epsilon_1\cdot \epsilon_3)(\epsilon_2\cdot \epsilon_4)
}_{\mathrm{(IV)}\,(A_1\cdot A_2)(A_3\cdot A_4)-(2\leftrightarrow3)}
\label{eq:y-three-point-longitudinal}\\
&\quad
+
\bigg[
\underbrace{
2(A_{12}\cdot p_3)(\epsilon_3\cdot \epsilon_4)
}_{\mathrm{(I)}\,2(A_{12}\cdot p_3)(A_3\cdot A_4)}
+
\underbrace{
(p_{12}\cdot A_{12})(\epsilon_3\cdot \epsilon_4)
}_{\mathrm{(II)}\,(p_{12}\cdot A_{12})(A_3\cdot A_4)}
-
\underbrace{
(A_{12}\cdot \epsilon_3)(p_3\cdot \epsilon_4)
}_{\mathrm{(III)}\,(A_{12}\cdot A_3)(p_3\cdot A_4)}
-(Q\leftrightarrow R)
\bigg]
\notag\\
&\quad
+
\underbrace{
(\epsilon_3\cdot \epsilon_1)(\epsilon_2\cdot \epsilon_4)
-(\epsilon_3\cdot \epsilon_2)(\epsilon_1\cdot \epsilon_4)
}_{\mathrm{(IV)}\,(A_3\cdot A_1)(A_2\cdot A_4)-(1\leftrightarrow2)}
\Bigg].
\notag
\end{align}
\endgroup
The first two lines and the next two lines correspond to the contributions from $(Q=1,R=23)$ and $(Q=12,R=3)$, respectively. The two-particle currents are
\begin{align*}
A_{23}^{\mu}
&=
\frac{1}{2s_{23}}
\Big[
\big(\epsilon_2\cdot(p_2+2p_3)\big)\epsilon_3^\mu
-
\big(\epsilon_3\cdot(p_3+2p_2)\big)\epsilon_2^\mu
+
(p_2-p_3)^\mu(\epsilon_2\cdot \epsilon_3)
\Big],
\\
A_{12}^{\mu}
&=
\frac{1}{2s_{12}}
\Big[
\big(\epsilon_1\cdot(p_1+2p_2)\big)\epsilon_2^\mu
-
\big(\epsilon_2\cdot(p_2+2p_1)\big)\epsilon_1^\mu
+
(p_1-p_2)^\mu(\epsilon_1\cdot \epsilon_2)
\Big].
\end{align*}
Applying $\cL_{13}^{\{1,2,3,4\}}\cT_{24}$ then gives
\begin{align}
\mathrm{(I)}\qquad
&\cL_{13}^{\{1,2,3,4\}}\cT_{24}
\Big[2(\epsilon_1\cdot p_{23})(A_{23}\cdot \epsilon_4)\Big]
=
-\frac{2(p_1\cdot p_{23})(p_2\cdot p_3)}{s_{23}},
\notag\\
\mathrm{(II)}\qquad
&\cL_{13}^{\{1,2,3,4\}}\cT_{24}
\Big[(p_1\cdot \epsilon_1)(A_{23}\cdot \epsilon_4)\Big]
=0,
\notag\\
\mathrm{(III)}\qquad
&\cL_{13}^{\{1,2,3,4\}}\cT_{24}
\Big[(\epsilon_1\cdot A_{23})(p_{23}\cdot \epsilon_4)\Big]
=0,
\notag\\
\mathrm{(IV)}\qquad
&\cL_{13}^{\{1,2,3,4\}}\cT_{24}
\Big[
(\epsilon_1\cdot \epsilon_2)(\epsilon_3\cdot \epsilon_4)
-(\epsilon_1\cdot \epsilon_3)(\epsilon_2\cdot \epsilon_4)
\Big]
=0,
\notag\displaybreak[3]\\[1mm]
\mathrm{(I)}\qquad
&\cL_{13}^{\{1,2,3,4\}}\cT_{24}
\Big[2(A_{12}\cdot p_3)(\epsilon_3\cdot \epsilon_4)\Big]
=0,
\notag\\
\mathrm{(II)}\qquad
&\cL_{13}^{\{1,2,3,4\}}\cT_{24}
\Big[(p_{12}\cdot A_{12})(\epsilon_3\cdot \epsilon_4)\Big]
=0,
\notag\\
\mathrm{(III)}\qquad
&\cL_{13}^{\{1,2,3,4\}}\cT_{24}
\Big[(A_{12}\cdot \epsilon_3)(p_3\cdot \epsilon_4)\Big]
=0,
\notag\\
\mathrm{(IV)}\qquad
&\cL_{13}^{\{1,2,3,4\}}\cT_{24}
\Big[
(\epsilon_3\cdot \epsilon_1)(\epsilon_2\cdot \epsilon_4)
-(\epsilon_3\cdot \epsilon_2)(\epsilon_1\cdot \epsilon_4)
\Big]
=0.
\label{eq:twop}
\end{align}
Substituting eq.~\eqref{eq:twop} and its $(Q\leftrightarrow R)$ terms into
eq.~\eqref{eq:y-three-point-longitudinal} gives
\begin{align}
\cL_{13}^{\{1,2,3,4\}}\cT_{24}
A_{123}\cdot \epsilon_4
=
-\frac{1}{s_{123}}\Big[\frac{(p_1\cdot p_{23})(p_2\cdot p_3)}{s_{23}}+
\frac{(p_3\cdot p_{12})(p_1\cdot p_2)}{s_{12}}\Big]
\nonumber.
\label{eq:y-three-point-check}
\end{align}
The right-hand side evaluates to
\begin{align*}
Y_{1\underline{2}3}
&=
\frac{1}{s_{123}}
\Big[
(p_1\cdot p_{23})Y_{\underline{2}3}
-
(p_3\cdot p_{12})Y_{1\underline{2}}
\Big]
\\
&=
-\frac{1}{s_{123}}
\left[
\frac{(p_1\cdot p_{23})(p_2\cdot p_3)}{s_{23}}
+
\frac{(p_3\cdot p_{12})(p_1\cdot p_2)}{s_{12}}
\right].
\end{align*}
Thus, we indeed have\[\cL_{13}^{\{1,2,3,4\}}\cT_{24}
A_{123}\cdot A_4=Y_{1\underline{2}3}Y_{\underline{4}}.\]

\subsection{The \texorpdfstring{$X$}{X}-current and the off-shell term}
\label{sec:x-current-relation}
One-loop NLSM amplitudes can be studied using only the relation in section~\ref*{sec:y-current-relation}; this subsection provides a more complete discussion of tree-level unifying relations. We label the two $Y$ external legs as $i,j$ in the order in which they appear in the word, so $i$ precedes $j$. Before proceeding, we state two lemmas that will be needed. Their proofs are lengthy, so the detailed proofs are given in appendices~\ref{eq:two-y-current-lemma-proof} and~\ref{eq:x-current-lemma-proof}, respectively.

\begin{lemma}
\label{lem:two-y-current}
Let \(M=QR\), with \(i\in Q\) and \(j\in R\). External legs \(i,j\) are \(Y\)-states, and all external legs in \(M\setminus\{i,j\}\) are \(Z\)-states. Then
\begin{equation}
\cL_{M\setminus\{i,j\}}^{M}\cT_{ij}\big(A_Q\cdot A_R\big)
=
Y_QY_R .
\label{eq:two-y-current-lemma}
\end{equation}
\end{lemma}

\begin{lemma}
\label{lem:x-current}
Let \(M=QR\), with \(i,j\in R\), and let all external legs in \(Q\) be \(Z\)-states. Then
\begin{equation}
\cL_{M\setminus\{i,j\}}^{M}\cT_{ij}\big(A_R\cdot A_Q\big)
=
Z_Q\cdot X_R+\frac{Z_Q\cdot p_R}{2s_R}\,\Gamma_R,
\label{eq:x-current-lemma}
\end{equation}
Here, $\Gamma_R$ is defined recursively, with the initial condition $
\Gamma_P=0\quad (|P|\leq2),$
\begin{equation}
\Gamma_R
:=
\sum_{\substack{R=ST\\ i,j\in T}}
\left(
Z_S\cdot X_T
+
\frac{Z_S\cdot p_T}{2s_T}\Gamma_T
\right)
-
\sum_{\substack{R=ST\\ i,j\in S}}
\left(
Z_T\cdot X_S
+
\frac{Z_T\cdot p_S}{2s_S}\Gamma_S
\right).
\label{eq:Gamma-recursion}
\end{equation}
\end{lemma}

We now consider the case in which both $Y$ external legs lie within the word $M$.
\begin{xrelation}
\phantomsection\label{lem:x-current-pn}
Let \(i,j\in M\), with external legs \(i,j\) being the two \(Y\)-state legs selected by the trace operator, and let the corresponding multiparticle current be \(X_M\), with \(n\notin M\). Then
\begin{equation}
\cL_{M\cup n\setminus\{i,j\}}^{M}\cT_{ij}
\big(A_M\cdot \epsilon_n\big)
=
X_M\cdot p_n
+
\frac{p_n\cdot p_M}{2s_M}\Gamma_M .
\label{eq:x-current-pn-lemma}
\end{equation}

Here, \(\Gamma_M\) satisfies the recursion~\eqref{eq:Gamma-recursion} defined above.
\end{xrelation}

We first establish the $X$-current relation for any word $N$ whose endpoints are $i$ and $j$. We then fix $N$ and proceed by induction on the word length $|M|$, with $|M|>|N|$. The case in which the two $Y$-legs lie at the endpoints of the word is a special case of the $X$-current relation. As can be seen from the proof below, when $i$ and $j$ are the endpoints of $M$, the terms arising from the deconcatenations do not contain expressions of the form appearing on the left-hand side of the $X$-current relation. Consequently, the induction hypothesis for the $X$-current relation itself is not needed, and the relation holds in this case. To avoid repeating the argument, we omit a separate proof of this case, which the reader can verify by following the steps below.

Assume that the $X$-current relation holds for all words containing $N$ as a contiguous subword and having length strictly between $|N|$ and $|M|$. Starting from the YM multiparticle-current recursion for \(A_M\) and contracting with \(\epsilon_n\) gives
\begin{equation}
\begin{aligned}
A_M\cdot \epsilon_n
=
\frac{1}{2s_M}
\sum_{M=ST}
\Bigg[
&\underbrace{
2(A_S\cdot p_T)(A_T\cdot \epsilon_n)
}_{\mathrm{(I)}}
+
\underbrace{
(p_S\cdot A_S)(A_T\cdot \epsilon_n)
}_{\mathrm{(II)}}
-
\underbrace{
(A_S\cdot A_T)(p_T\cdot \epsilon_n)
}_{\mathrm{(III)}}
\\
&-
\sum_{T=UV}
\Big(
\underbrace{
(A_S\cdot A_U)(A_V\cdot \epsilon_n)
}_{\mathrm{(IV)}}
-(U\leftrightarrow V)
\Big)
-(S\leftrightarrow T)
\Bigg].
\end{aligned}
\label{eq:x-pn-bg}
\end{equation}
Apply \(\cL_{M\cup n\setminus\{i,j\}}^{M}\cT_{ij}\) to both sides of eq.~\eqref{eq:x-pn-bg}. The preceding discussion shows that \(\mathrm{(II)}\) and the four-current term \(\mathrm{(IV)}\) give no contribution under this operator. Thus, only \(\mathrm{(I)}\), \(\mathrm{(III)}\), and their \(S\leftrightarrow T\) terms need to be considered. According to the positions of \(i,j\) in the split \(M=ST\), there are
\[
i,j\in S,\qquad i\in S,\ j\in T,\qquad i,j\in T
\]
three cases.

First consider \(i,j\in S\). For the unexchanged term in \(\mathrm{(I)}\),
\[
\begin{aligned}
\cL_{M\cup n\setminus\{i,j\}}^{M}\cT_{ij}
\left[(A_S\cdot p_T)(A_T\cdot \epsilon_n)\right]
=
\cL_{S\setminus\{i,j\}}^{M}\cT_{ij}(A_S\cdot p_T)\,
\cL_{T\cup n}^{T}(A_T\cdot \epsilon_n)
=0.
\end{aligned}
\]
Thus, only the term obtained after \(S\leftrightarrow T\),
$(A_T\cdot p_S)(A_S\cdot \epsilon_n)$
contributes. Since \(T\) is a pure-\(Z\) subword, the relation in section~\ref*{sec:z-current-relation} and the induction hypothesis give
\[
\begin{aligned}
&\cL_{M\cup n\setminus\{i,j\}}^{M}\cT_{ij}
\left[(A_T\cdot p_S)(A_S\cdot \epsilon_n)\right]
=
(Z_T\cdot p_S)
\left(
X_S\cdot p_n
+
\frac{p_n\cdot p_S}{2s_S}\Gamma_S
\right).
\end{aligned}
\]
For \(\mathrm{(III)}\), lemma~\ref{lem:x-current} gives
\[
\cL_{M\setminus\{i,j\}}^{M}\cT_{ij}(A_S\cdot A_T)
=
X_S\cdot Z_T
+
\frac{Z_T\cdot p_S}{2s_S}\Gamma_S.
\]
Using, in addition,
\[
\cL_n^T(p_T\cdot \epsilon_n)=p_T\cdot p_n,
\qquad
\cL_n^S(p_S\cdot \epsilon_n)=p_S\cdot p_n,
\]
the total contribution of this term is
\begingroup\small
\begin{align}
\frac{1}{2s_M}
\sum_{\substack{M=ST\\ i,j\in S}}
\Bigg[
-2(Z_T\cdot p_S)
\left(
X_S\cdot p_n
+\frac{p_n\cdot p_S}{2s_S}\Gamma_S
\right)
-
\left(
X_S\cdot Z_T
+\frac{Z_T\cdot p_S}{2s_S}\Gamma_S
\right)
\left(
p_T\cdot p_n-p_S\cdot p_n
\right)
\Bigg].
\label{eq:x-pn-sector-S}
\end{align}
\endgroup

Next consider \(i\in S,\ j\in T\). In this case, \(\epsilon_i,\epsilon_j\) cannot form \(\epsilon_i\cdot \epsilon_j\) in \(\mathrm{(I)}\) or its exchanged term, so these terms are directly annihilated by \(\cT_{ij}\). For \(\mathrm{(III)}\), lemma~\ref{lem:two-y-current} gives
\[
\cL_{M\setminus\{i,j\}}^{M}\cT_{ij}(A_S\cdot A_T)
=
Y_SY_T.
\]
Hence, the contribution from this part is
\begin{equation}
\frac{1}{2s_M}
\sum_{\substack{M=ST\\ i\in S,\ j\in T}}
Y_SY_T
\left(
p_S\cdot p_n-p_T\cdot p_n
\right).
\label{eq:x-pn-sector-ST}
\end{equation}

Finally, consider \(i,j\in T\). This case is symmetric to \(i,j\in S\), so its contribution is
\begingroup\small
\begin{align}
\frac{1}{2s_M}
\sum_{\substack{M=ST\\ i,j\in T}}
\Bigg[
2(Z_S\cdot p_T)
\left(
X_T\cdot p_n
+\frac{p_n\cdot p_T}{2s_T}\Gamma_T
\right)
-
\left(
X_T\cdot Z_S
+\frac{Z_S\cdot p_T}{2s_T}\Gamma_T
\right)
\left(
p_T\cdot p_n-p_S\cdot p_n
\right)
\Bigg].\nonumber
\label{eq:x-pn-sector-T}
\end{align}
\endgroup

Adding the contributions from the three types of splits gives
\begingroup\small
\begin{equation}
\begin{aligned}
&\cL_{M\cup n\setminus\{i,j\}}^{M}\cT_{ij}(A_M\cdot \epsilon_n)
\\
&={}
\frac{1}{2s_M}
\Bigg[
\sum_{\substack{M=ST\\ i,j\in S}}
\Bigg\{
-
\left(
X_S\cdot Z_T
+\frac{Z_T\cdot p_S}{2s_S}\Gamma_S
\right)
\left(
p_T\cdot p_n-p_S\cdot p_n
\right)
\\
&-2(Z_T\cdot p_S)
\left(
X_S\cdot p_n
+\frac{p_n\cdot p_S}{2s_S}\Gamma_S
\right)
\Bigg\}
+
\sum_{\substack{M=ST\\ i\in S,\ j\in T}}
Y_SY_T
\left(
p_S\cdot p_n-p_T\cdot p_n
\right)
\\
&+\sum_{\substack{M=ST\\ i,j\in T}}
\Bigg\{
2(Z_S\cdot p_T)
\left(
X_T\cdot p_n
+\frac{p_n\cdot p_T}{2s_T}\Gamma_T
\right)
-
\left(
X_T\cdot Z_S
+\frac{Z_S\cdot p_T}{2s_T}\Gamma_T
\right)
\left(
p_T\cdot p_n-p_S\cdot p_n
\right)
\Bigg\}
\Bigg].
\end{aligned}
\label{eq:x-pn-lhs-expanded}
\end{equation}
\endgroup
We now turn to the NLSM side. The multiparticle-current recursion for the NLSM \(X\)-current gives
\begin{align}
X_M\cdot p_n
=
-\frac{1}{s_M}
\sum_{M=ST}
\Bigg[
(p_T\cdot p_n)
\left(
\frac12Y_SY_T+Z_S\cdot X_T
\right)
+
(p_S\cdot Z_T)(X_S\cdot p_n)
-(S\leftrightarrow T)
\Bigg].\nonumber
\label{eq:XM-pn}
\end{align}
We again classify the splits into the three types above. If \(i,j\in S\), lemmas~\ref{lem:non-z-z-zero} and~\ref{lem:even-non-z-y-zero} imply
\[
Y_SY_T=0,\qquad
Z_S\cdot X_T=0,\qquad
p_T\cdot Z_S=0,
\]
so this sector contributes
\[
\frac{1}{2s_M}
\sum_{\substack{M=ST\\ i,j\in S}}
\left[
-2(Z_T\cdot p_S)(X_S\cdot p_n)
+2(p_S\cdot p_n)(X_S\cdot Z_T)
\right].
\]
If \(i\in S,\ j\in T\), lemma~\ref{lem:non-z-z-zero} implies that all terms containing \(Z\) vanish, leaving only
\[
\frac{1}{2s_M}
\sum_{\substack{M=ST\\ i\in S,\ j\in T}}
Y_SY_T
\left(
p_S\cdot p_n-p_T\cdot p_n
\right).
\]
If \(i,j\in T\), the same reasoning as for \(i,j\in S\) gives
\[
\frac{1}{2s_M}
\sum_{\substack{M=ST\\ i,j\in T}}
\left[
-2(p_T\cdot p_n)(Z_S\cdot X_T)
+2(Z_S\cdot p_T)(X_T\cdot p_n)
\right].
\]

Subtracting the sum of these three contributions from eq.~\eqref{eq:x-pn-lhs-expanded} gives
\begin{align}
\frac{p_M\cdot p_n}{2s_M}
\Bigg[
\sum_{\substack{M=ST\\ i,j\in T}}
\left(
X_T\cdot Z_S
+
\frac{Z_S\cdot p_T}{2s_T}\Gamma_T
\right)
-
\sum_{\substack{M=ST\\ i,j\in S}}
\left(
X_S\cdot Z_T
+
\frac{Z_T\cdot p_S}{2s_S}\Gamma_S
\right)
\Bigg].
\end{align}
By the recursion~\eqref{eq:Gamma-recursion} for \(\Gamma_M\), the quantity in square brackets is precisely \(\Gamma_M\). Hence,
\[
\cL_{M\cup n\setminus\{i,j\}}^{M}\cT_{ij}(A_M\cdot A_n)
=
X_M\cdot Z_n
+
\frac{p_M\cdot p_n}{2s_M}\Gamma_M.
\]
This completes the induction and proves the relation in section~\ref*{sec:x-current-relation}.

Finally, consider the special case in which \(i,j\) lie at the two endpoints of the word \(M\). For any binary split \(M=ST\), neither $i,j\in S$ nor $
i,j\in T.$ is possible. The only possible type is $i\in S$ and $j\in T.$ Thus, both sums in the recursion for \(\Gamma_M\) are empty, and $\Gamma_M=0$. This is consistent with the fact that the proof for such words does not rely on the induction hypothesis for the $X$-current relation.
In this case, the relation in section~\ref*{sec:x-current-relation} therefore reduces to
\begin{equation}
\cL_{M\cup n\setminus\{i,j\}}^{M}\cT_{ij}(A_M\cdot A_n)
=
X_M\cdot Z_n.
\label{eq:x-current-endpoint}
\end{equation}

We give two further examples of multiparticle currents with word length $3$ to verify the relation in section~\ref*{sec:x-current-relation} and its special case, eq.~\eqref{eq:x-current-endpoint}.

First take
$
M=123,\quad n=4,\quad ij=23.
$
Applying
$\cL_{14}^{\{1,2,3\}}\cT_{23}$ to each term in eq.~\eqref{eq:y-three-point-longitudinal} gives the nonzero terms
\begin{align*}
\mathrm{(I)}\qquad
&\cL_{14}^{\{1,2,3\}}\cT_{23}
\Big[2(\epsilon_1\cdot p_{23})(A_{23}\cdot \epsilon_4)\Big]
=
\frac{(p_1\cdot p_{23})
\big(p_2\cdot p_4-p_3\cdot p_4\big)}{s_{23}},
\\
\mathrm{(III)}\qquad
&\cL_{14}^{\{1,2,3\}}\cT_{23}
\Big[(\epsilon_1\cdot A_{23})(p_{23}\cdot \epsilon_4)\Big]
=
\frac{(p_1\cdot p_2-p_1\cdot p_3)
(p_{23}\cdot p_4)}{2s_{23}},
\\
\mathrm{(III)}\qquad
&\cL_{14}^{\{1,2,3\}}\cT_{23}
\Big[(A_{12}\cdot \epsilon_3)(p_3\cdot \epsilon_4)\Big]
=
\frac{p_1\cdot p_2}{s_{12}}\,
p_3\cdot p_4.
\end{align*}
The nonzero terms among the $(Q\leftrightarrow R)$ terms are
$
\frac{(p_1\cdot p_2-p_1\cdot p_3)(p_1\cdot p_4)}
{2s_{23}},
$
and
$
\frac{p_1\cdot p_2}{s_{12}}\,
p_{12}\cdot p_4
$.
Substituting back into eq.~\eqref{eq:y-three-point-longitudinal} gives
\begin{align}
&\cL_{14}^{\{1,2,3\}}\cT_{23}
\big(A_{123}\cdot \epsilon_4\big)
\nonumber\\
={}&
\frac{1}{2s_{123}}
\Bigg[
\frac{(p_1\cdot p_{23})
\big(p_2\cdot p_4-p_3\cdot p_4\big)}{s_{23}}
-
\frac{(p_1\cdot p_2-p_1\cdot p_3)
(p_{23}\cdot p_4)}{2s_{23}}
\nonumber\\
&\hspace{2.2cm}
+
\frac{(p_1\cdot p_2-p_1\cdot p_3)
(p_1\cdot p_4)}{2s_{23}}
-
\frac{p_1\cdot p_2}{s_{12}}\,
p_3\cdot p_4
+
\frac{p_1\cdot p_2}{s_{12}}\,
p_{12}\cdot p_4
\Bigg].
\label{eq:x-three-point-check-23}
\end{align}

We now evaluate the right-hand side. External legs $2,3$ are $Y$-states, while leg $1$ is a $Z$-state. First,
\[
Z_1^\mu=p_1^\mu,
\qquad
X_{23}^\mu
=
-\frac{1}{2s_{23}}
(p_3^\mu-p_2^\mu),
\qquad
Y_{12}=\frac{p_1\cdot p_2}{s_{12}}.
\]
Thus,
\begin{align}
\Gamma_{123}
=
Z_1\cdot X_{23}
=
\frac{p_1\cdot p_2-p_1\cdot p_3}{2s_{23}}.
\label{eq:Gamma123-23}
\end{align}
The $X$-current recursion gives
\begin{align*}
X_{123}\cdot p_4
&=
-\frac{1}{s_{123}}
\Bigg[
p_{23}\cdot p_4(Z_1\cdot X_{23})
-(p_{23}\cdot Z_1)X_{23}\cdot p_4
+\frac12(p_3-p_{12})\cdot p_4\,Y_{12}
\Bigg]
\\
&=
\frac{
(p_1\cdot p_3)(p_2\cdot p_4)
-(p_1\cdot p_2)(p_3\cdot p_4)
}{
s_{123}s_{23}
}
+
\frac{
p_1\cdot p_2
}{
2s_{123}s_{12}
}
\Big(
p_1\cdot p_4
+p_2\cdot p_4
-p_3\cdot p_4
\Big).
\end{align*}
while
\[
\frac{p_{123}\cdot p_4}{2s_{123}}\Gamma_{123}
=
\frac{
(p_{123}\cdot p_4)
(p_1\cdot p_2-p_1\cdot p_3)
}{
4s_{123}s_{23}
}.
\]
Thus,
\begin{equation}
\begin{aligned}
&X_{123}\cdot p_4
+
\frac{p_{123}\cdot p_4}{2s_{123}}\Gamma_{123}
\\
={}&
\frac{
(p_1\cdot p_3)(p_2\cdot p_4)
-(p_1\cdot p_2)(p_3\cdot p_4)
}{
s_{123}s_{23}
}
+
\frac{
p_1\cdot p_2
}{
2s_{123}s_{12}
}
\Big(
p_1\cdot p_4
+p_2\cdot p_4
-p_3\cdot p_4
\Big)
\\
&\qquad+
\frac{
(p_{123}\cdot p_4)
(p_1\cdot p_2-p_1\cdot p_3)
}{
4s_{123}s_{23}
}.
\end{aligned}
\label{eq:x-three-point-rhs-23}
\end{equation}
Substituting $p_{123}=p_1+p_2+p_3$ and simplifying shows that
eq.~\eqref{eq:x-three-point-rhs-23} and
eq.~\eqref{eq:x-three-point-check-23} are exactly equal.

Next take
$
M=123,\quad n=4,\quad ij=13.
$
Applying
$\cL_{24}^{\{1,2,3\}}\cT_{13}$ to each explicitly displayed term in eq.~\eqref{eq:y-three-point-longitudinal} gives the nonzero terms
\begin{align*}
\mathrm{(III)}\qquad
&\cL_{24}^{\{1,2,3\}}\cT_{13}
\Big[(\epsilon_1\cdot A_{23})(p_{23}\cdot \epsilon_4)\Big]
=
\frac{p_2\cdot p_3}{s_{23}}\,
p_{23}\cdot p_4,
\\
\mathrm{(III)}\qquad
&\cL_{24}^{\{1,2,3\}}\cT_{13}
\Big[(A_{12}\cdot \epsilon_3)(p_3\cdot \epsilon_4)\Big]
=
-\frac{p_1\cdot p_2}{s_{12}}\,
p_3\cdot p_4.
\end{align*}
The corresponding $(Q\leftrightarrow R)$ terms give, respectively,
$
\frac{p_2\cdot p_3}{s_{23}}\,
p_1\cdot p_4,
$
and
$
-\frac{p_1\cdot p_2}{s_{12}}\,
p_{12}\cdot p_4,
$
The remaining exchanged terms still vanish. Thus,
\[
\cL_{24}^{\{1,2,3\}}\cT_{13}
\big(A_{123}\cdot \epsilon_4\big)
=
\frac{1}{2s_{123}}
\Bigg[
-\frac{p_2\cdot p_3}{s_{23}}\,
p_{23}\cdot p_4
+\frac{p_2\cdot p_3}{s_{23}}\,
p_1\cdot p_4
+\frac{p_1\cdot p_2}{s_{12}}\,
p_3\cdot p_4
-\frac{p_1\cdot p_2}{s_{12}}\,
p_{12}\cdot p_4
\Bigg].
\]

On the NLSM side, external legs $1,3$ are $Y$-states, while leg $2$ is a $Z$-state. Thus,
\[
Y_{23}
=
\frac{p_2\cdot p_3}{s_{23}},
\qquad
Y_{12}
=
-\frac{p_1\cdot p_2}{s_{12}}.
\]
The $X$-current recursion gives
\begin{align*}
X_{123}^{\mu}
&=
-\frac{1}{2s_{123}}
\Big[
(p_{23}-p_1)^\mu Y_{23}
+
(p_3-p_{12})^\mu Y_{12}
\Big]
\\
&=
-\frac{1}{2s_{123}}
\Bigg[
\frac{p_2\cdot p_3}{s_{23}}
(p_{23}-p_1)^\mu
-
\frac{p_1\cdot p_2}{s_{12}}
(p_3-p_{12})^\mu
\Bigg],
\end{align*}
and hence
\[
X_{123}\cdot p_4
=
\frac{1}{2s_{123}}
\Bigg[
-\frac{p_2\cdot p_3}{s_{23}}\,
p_{23}\cdot p_4
+\frac{p_2\cdot p_3}{s_{23}}\,
p_1\cdot p_4
+\frac{p_1\cdot p_2}{s_{12}}\,
p_3\cdot p_4
-\frac{p_1\cdot p_2}{s_{12}}\,
p_{12}\cdot p_4
\Bigg].
\]
On the other hand, since $1$ and $3$ are the two endpoints of the word $123$, no binary split has
$1,3\in S$ or $1,3\in T$, so
\[
\Gamma_{123}=0.
\]
Thus,
\[
X_{123}\cdot p_4
+
\frac{p_{123}\cdot p_4}{2s_{123}}\Gamma_{123}
=
\frac{1}{2s_{123}}
\Bigg[
-\frac{p_2\cdot p_3}{s_{23}}\,
p_{23}\cdot p_4
+\frac{p_2\cdot p_3}{s_{23}}\,
p_1\cdot p_4
+\frac{p_1\cdot p_2}{s_{12}}\,
p_3\cdot p_4
-\frac{p_1\cdot p_2}{s_{12}}\,
p_{12}\cdot p_4
\Bigg],
\]
which agrees exactly with the result of evaluating
$\cL_{24}^{\{1,2,3\}}\cT_{13}
\big(A_{123}\cdot \epsilon_4\big)$
.

Thus, the relation in section~\ref*{sec:x-current-relation} is verified in both examples of multiparticle currents with word length $3$.

\paragraph{Discussion.} We can now discuss the unifying relation between the Feynman diagrams of the two theories.
We have established that any YM Feynman diagram containing a four-point vertex vanishes under the transmutation operator. YM Feynman diagrams giving nonzero results after transmutation can therefore contain only three-point vertices. The following method then identifies, term by term, the corresponding YM and NLSM Feynman diagrams.

In the derivations and explicit calculations above, propagator factors such as $\frac{1}{2s_M}$ are never required to be further reduced, and the transmutation operators themselves do not act on these factors. We can therefore regard the different $s_M^{-1}$ as mutually independent formal variables and compare the two sides term by term in propagator monomials. For a given propagator monomial, its coefficient is precisely the contribution of the corresponding Feynman diagram with those propagator factors stripped off. For example, the monomial
\[
\frac{1}{s_{123}s_{23}}
\]
corresponds to the following propagator structure: external legs $2,3$ first join to form the internal line $p_{23}$, which then joins external leg $1$ to form the internal line $p_{123}$, and finally connects to external leg $4$. Thus, in the off-shell construction considered here, the unifying relation for color-ordered amplitudes can be refined to hold term by term in propagator structures, that is, at the level of the corresponding tree-level Feynman diagrams; the corresponding diagrams even have identical topologies.
\section{One-loop unifying relations}
\label{sec:sewing-definitions}
The discussion above of the unifying relations between YM and NLSM multiparticle currents suggests that their one-loop Feynman diagrams can also be connected through a transmutation operator. Taking one-loop integrands as the objects of interest, we call the operator implementing the unifying relation between one-loop integrands a one-loop operator and denote it by $O$.

In the multiparticle-current construction, a one-loop integrand can be obtained by sewing two off-shell external legs of a tree-level object into an internal line
\cite{Gomez:2022dzk,Lee:2022aiu,Chen:2023dcx}. This means that the tree-level unifying relation can be used to construct the one-loop operator. Denoting the sewing operation by $\cS$, we require that applying the one-loop operator to the sewn YM tree-level object give the same result, in the amplitude limit, as first applying the tree-level operator to the tree-level object and then sewing to obtain the one-loop integrand. Clearly, it suffices to establish
\[
O\,\cS =\cS \,\cL\!\cdot\cT.
\]

For YM theory, let $i,j$ be the two external legs to be sewn. Sewing replaces $\epsilon_i^\mu\epsilon_j^\nu$ in each monomial with $\eta^{\mu\nu}$ and then assigns equal and opposite momenta to the two legs. Denoting the momentum of the resulting internal line by $\ell$, we set $p_i=\ell,\,p_j=-\ell$ and finally multiply by the propagator $1/\ell^2$ of the internal line formed by sewing. This operation can be written as
\begin{equation}
\epsilon_i^\mu\,\epsilon_j^\nu\rightarrow \frac{\eta^{\mu\nu}}{\ell^2},
\label{eq:vector-sewing}
\end{equation}
Whenever $\eta^{\mu\nu}\eta_{\mu\nu}$ appears during sewing, we treat it as a new formal variable, denoted by $d$, which is the spacetime dimension of the theory.
The trace operator $\cT_{ij}$ extracts the coefficient of $\epsilon_i\cdot \epsilon_j$ in the tree-level object. After sewing, $\epsilon_i\cdot \epsilon_j$ becomes $d$, so\[
\frac{\partial}{\partial d}\,\cS=\cS\,\cT_{ij}.\]
This gives one component of the one-loop operator, $\frac{\partial}{\partial d}$, denoted by $D$. The operator $D$ differentiates only the explicit dependence of the integrand on the formal Lorentz-trace counting variable $d$, with all other variables held fixed, and acts before loop integration.

For the NLSM, the two legs to be sewn are $Y$ legs, so no Lorentz indices need to be contracted. The remaining steps are the same as for sewing gluon external legs; thus, the corresponding sewing operation only requires setting $p_i=\ell,\,p_j=-\ell.$

In Feynman-gauge YM theory, loop diagrams must also include ghost contributions. The one-loop ghost integrand is obtained from tree diagrams by sewing two ghost external states. Consequently, it contains no $d$ after sewing and is annihilated by $D$. Thus, only gluon-loop contributions need to be considered in the one-loop unifying relation.

We next discuss the operator into which a longitudinal operator $\cL_r,\,r\neq i,j$ is converted when commuted past sewing.
During sewing, the assignment of momenta to the two sewn legs can affect $\cL_r$. The potentially affected terms in $\cL_r$ involve $\epsilon_r\!\cdot p_i$ and $\epsilon_r\!\cdot p_j$. Sewing converts $\epsilon_r\!\cdot p_i$ and $\epsilon_r\!\cdot p_j$ into $\epsilon_r\!\cdot \ell$ and $-\epsilon_r\!\cdot \ell$, respectively. If the longitudinal operator acts first, $\epsilon_r\!\cdot p_i$ and $\epsilon_r\!\cdot p_j$ are replaced by $p_r\!\cdot p_i$ and $p_r\!\cdot p_j$, which become $p_r\!\cdot \ell$ and $-p_r\!\cdot \ell$ after sewing. A direct comparison gives the correspondence between the tree-level operator
\[
(p_r\!\cdot p_i)
\frac{\partial}
{\partial(\epsilon_r\!\cdot p_i)}
+
(p_r\!\cdot p_j)
\frac{\partial}
{\partial(\epsilon_r\!\cdot p_j)}
\]
and the one-loop operator
\[
(p_r\!\cdot \ell)
\frac{\partial}
{\partial(\epsilon_r\!\cdot \ell)}.
\]
The remaining terms in the longitudinal operator, which do not involve $i,j$, commute with sewing. We therefore obtain
\begin{equation}
\cL_r^{\ell;M\setminus\{i,j\}}\,
\cS
=
\cS
\cL_r^{M},
\label{eq:longitudinal-sewing-commutation}
\end{equation}
where
\begin{equation}
\cL_r^{\ell;M\setminus\{i,j\}}
=
(p_r\!\cdot\ell)
\frac{\partial}
{\partial(\epsilon_r\!\cdot\ell)}
+
\sum_{\substack{q\in M\setminus\{i,j\}\\q\neq r}}
(p_r\!\cdot p_q)
\frac{\partial}
{\partial(\epsilon_r\!\cdot p_q)}.
\end{equation}

Combining this with the relation between $D$ and $\cT_{ij}$ gives
\begin{align}
\cL^{\ell;M\setminus\{i,j\}}_{M\setminus\{i,j\}}
D\,\cS=
\cS
\cL_{M\setminus\{i,j\}}^{M}
\cT_{ij},
\label{eq:complete-sewing-commutation}
\end{align}
where
\[\cL^{\ell;M\setminus\{i,j\}}_{M\setminus\{i,j\}}:=
\prod_{r\in M\setminus\{i,j\}}
\cL_r^{\ell;M\setminus\{i,j\}}.
\]
$\cL^{\ell;M\setminus\{i,j\}}_{M\setminus\{i,j\}}
D$ is the one-loop operator $O$ that we have obtained.

The one-loop operator in eq.~\eqref{eq:complete-sewing-commutation} has the same differential structure as $\mathcal L_\circ^\epsilon D$ (in the present notation) used in the CHY forward-limit representation of ref.~\cite{Zhou:2021gnu}: it first extracts the trace over internal loop states and then applies longitudinal operators to the external legs.

Equation~\eqref{eq:complete-sewing-commutation} relates tree-level transmutation to one-loop integrands. Taking $M=lPn$ and sewing the auxiliary legs $l,n$ produces a one-loop diagram with color ordering $P$. Below, we take $P=123$ and follow the preceding diagram-by-diagram comparison to examine the correspondence explicitly. The one-loop transmutation operator is
\begin{equation}
O=\cL_1^{\ell;123}\cL_2^{\ell;123}\cL_3^{\ell;123}D.
\label{eq:ch4-example-operator}
\end{equation}

We retain the momentum and multiparticle-current notation used above, with all physical external momenta $p_i$ taken to be outgoing. Arrows next to propagators indicate the direction of the labeled loop momentum. For example, at the vertex carrying external leg $1$ in the triangle diagram, $\ell+p_1$ is incoming, whereas $\ell$ and $p_1$ are outgoing, so that $(\ell+p_1)-\ell-p_1=0$. Arrows on mixed $XZ$ propagators instead indicate the connection from the $X$ end to the $Z$ end. As explained above for ghost-loop terms, we need only consider gluon loops and the corresponding NLSM diagrams below.

\flushbottom
\begingroup
\setlength{\intextsep}{4pt plus 2pt minus 1pt}
\setlength{\abovecaptionskip}{5pt}
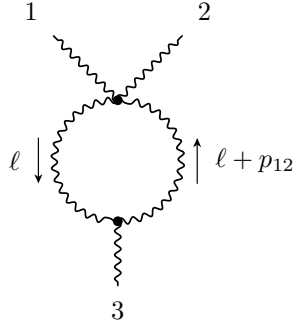
\begin{figure}[H]
\centering
\begin{tikzpicture}[chfour figure]

\coordinate (a) at (0,.8);
\coordinate (b) at (0,-.8);
\draw[chfour gluon] (a) arc[start angle=90,end angle=270,radius=.8];
\draw[chfour gluon] (b) arc[start angle=270,end angle=450,radius=.8];
\draw[chfour gluon] (a)--(-.85,1.65) node[above left=2pt] {$1$};
\draw[chfour gluon] (a)--(.85,1.65) node[above right=2pt] {$2$};
\draw[chfour gluon] (b)--(0,-1.65) node[below=2pt] {$3$};
\foreach \v in {a,b} \node[chfour vertex] at (\v) {};
\draw[chfour momentum] (-1.05,.3)--(-1.05,-.3) node[midway,left=3pt] {$\ell$};
\draw[chfour momentum] (1.05,-.3)--(1.05,.3) node[midway,right=3pt] {$\ell+p_{12}$};

\end{tikzpicture}
\caption{Bubble diagram containing a four-gluon vertex. External legs $1,2$ attach to the upper four-point vertex, and external leg $3$ to the lower three-point vertex. Wavy lines denote gluons, and arrows next to propagators indicate the direction of the loop momentum.}
\label{fig:ch4r-quartic}
\end{figure}
\endgroup

\paragraph{Bubble with a four-point interaction.}
The YM integrand contribution corresponding to figure~\ref{fig:ch4r-quartic} is
\begin{equation}
\begin{aligned}
I_{\mathrm{YM}}^{(4)}={} &\frac{1}{\ell^2(\ell+p_{12})^2}\\
&\times\Big[-\big((2\ell+2p_{12}+p_3)\cdot\varepsilon_3\big)\delta^{\mu}{}_{\nu}+\varepsilon_3^{\mu}(\ell+p_{12}+2p_3)_{\nu}
 +(\ell+p_{12}-p_3)^{\mu}\varepsilon_{3\nu}\Big]\\
&\times\Big[2\varepsilon_1^{\nu}\varepsilon_{2\mu}
 -\varepsilon_2^{\nu}\varepsilon_{1\mu}
 -(\varepsilon_1\cdot\varepsilon_2)\delta^{\nu}{}_{\mu}\Big].
\end{aligned}
\label{eq:ch4-quartic-integrand}
\end{equation}
The cubic $XYZ$ representation of the NLSM has no four-point vertex, so the corresponding contribution is
\begin{equation}
I_{\mathrm{NLSM}}^{(4)}=0.
\label{eq:ch4-quartic-NLSM}
\end{equation}
Applying the transmutation operator gives
\begin{equation}
OI_{\mathrm{YM}}^{(4)}=0=I_{\mathrm{NLSM}}^{(4)}.
\label{eq:ch4-quartic-relation}
\end{equation}

\begin{figure}[H]
\centering
\begin{tikzpicture}[chfour figure]
\begin{scope}

\coordinate (a) at (0,0);
\coordinate (b) at (1.1,0);
\coordinate (c) at (1.7,-.85);
\draw[chfour gluon] (a) arc[start angle=0,end angle=360,radius=.8];
\draw[chfour gluon] (a)--(b);
\draw[chfour gluon] (b)--(c);
\draw[chfour gluon] (b)--(1.65,.9) node[above right=2pt] {$3$};
\draw[chfour gluon] (c)--(1.42,-1.7) node[below=2pt] {$1$};
\draw[chfour gluon] (c)--(2.55,-1.3) node[right=2pt] {$2$};
\node[above=3pt] at (.55,0) {$p_{123}$};
\node[right=3pt] at (1.47,-.43) {$p_{12}$};
\foreach \v in {a,b,c} \node[chfour vertex] at (\v) {};
\draw[chfour momentum] (-1.61,.81) arc[start angle=135,end angle=225,radius=1.15];
\node[left=2pt] at (-1.98,0) {$\ell$};
\node at (.25,-2.12) {$\mathrm{YM}$};
\end{scope}
\begin{scope}[xshift=6.0cm]

\coordinate (a) at (0,0);
\coordinate (b) at (1.1,0);
\coordinate (c) at (1.7,-.85);
\draw[chfour scalar] (a) arc[start angle=0,end angle=360,radius=.8];
\draw[chfour mixed] (b)--(a);
\draw[chfour mixed] (c)--(b);
\draw[chfour external] (b)--(1.65,.9) node[above right=2pt] {$3$};
\draw[chfour external] (c)--(1.42,-1.7) node[below=2pt] {$1$};
\draw[chfour external] (c)--(2.55,-1.3) node[right=2pt] {$2$};
\node[above=3pt] at (.55,0) {$p_{123}$};
\node[right=3pt] at (1.47,-.43) {$p_{12}$};
\foreach \v in {a,b,c} \node[chfour vertex] at (\v) {};
\draw[chfour momentum] (-1.61,.81) arc[start angle=135,end angle=225,radius=1.15];
\node[left=2pt] at (-1.98,0) {$\ell$};
\node at (.25,-2.12) {$\mathrm{NLSM}$};
\end{scope}
\end{tikzpicture}
\caption{Tadpole diagrams. The left diagram contains a gluon loop. In the right diagram, the solid loop line denotes $Y$, the external dashed lines are $Z$ legs, and the internal dashed lines with arrows are mixed $XZ$ propagators, with arrows pointing from the $X$ end to the $Z$ end. The curved arrow outside each loop indicates only the direction of the momentum $\ell$.}
\label{fig:ch4r-tadpole}
\end{figure}
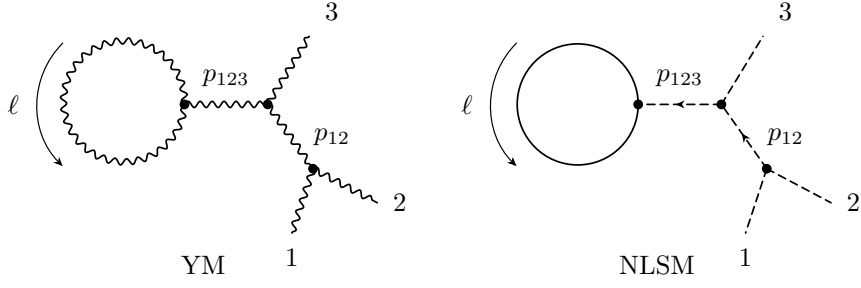

\paragraph{Tadpole diagram.}
The integrand contributions of the two diagrams in figure~\ref{fig:ch4r-tadpole} are
\begin{align}
I_{\mathrm{YM}}^{\mathrm{tad}}
&=-\frac{d-1}{\ell^2}(2\ell+p_{123})\cdot A_{(12)3},
\label{eq:ch4-tadpole-YM}\\
I_{\mathrm{NLSM}}^{\mathrm{tad}}
&=-\frac{2\ell\cdot Z_{(12)3}}{\ell^2}.
\label{eq:ch4-tadpole-NLSM}
\end{align}
Here $A_{(12)3}$ and $Z_{(12)3}$ denote the contributions associated with the attached tree subdiagram.
Applying the transmutation operator gives
\begin{equation}
OI_{\mathrm{YM}}^{\mathrm{tad}}
=-\frac{2\ell\cdot Z_{(12)3}}{\ell^2}
=I_{\mathrm{NLSM}}^{\mathrm{tad}}.
\label{eq:ch4-tadpole-relation}
\end{equation}
Thus, acting with the differential operator $O$ on the YM integrand reproduces the result obtained from a direct calculation of the NLSM tadpole diagram. After restoring the color factor, this diagram does not contribute to the full amplitude because $f^{abb}=0$. In a massless theory, this scaleless loop integral also vanishes in dimensional regularization.

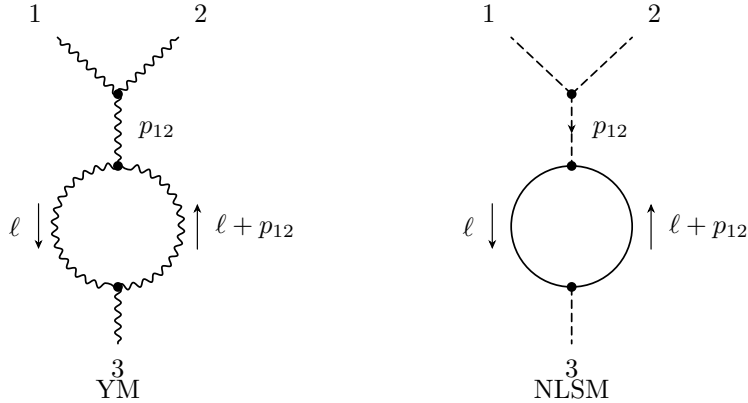
\begin{figure}[H]
\centering
\begin{tikzpicture}[chfour figure]
\begin{scope}

\coordinate (a) at (0,0);
\coordinate (b) at (0,-1.6);
\coordinate (c) at (0,.95);
\draw[chfour gluon] (a) arc[start angle=90,end angle=270,radius=.8];
\draw[chfour gluon] (b) arc[start angle=270,end angle=450,radius=.8];
\draw[chfour gluon] (c)--(a);
\node[right=4pt] at (0,.48) {$p_{12}$};
\draw[chfour gluon] (c)--(-.8,1.7) node[above left=2pt] {$1$};
\draw[chfour gluon] (c)--(.8,1.7) node[above right=2pt] {$2$};
\draw[chfour gluon] (b)--(0,-2.35) node[below=2pt] {$3$};
\foreach \v in {a,b,c} \node[chfour vertex] at (\v) {};
\draw[chfour momentum] (-1.05,-.5)--(-1.05,-1.1) node[midway,left=3pt] {$\ell$};
\draw[chfour momentum] (1.05,-1.1)--(1.05,-.5) node[midway,right=3pt] {$\ell+p_{12}$};
\node at (0,-2.95) {$\mathrm{YM}$};
\end{scope}
\begin{scope}[xshift=6.0cm]

\coordinate (a) at (0,0);
\coordinate (b) at (0,-1.6);
\coordinate (c) at (0,.95);
\draw[chfour scalar] (a) arc[start angle=90,end angle=270,radius=.8];
\draw[chfour scalar] (b) arc[start angle=270,end angle=450,radius=.8];
\draw[chfour mixed] (c)--(a);
\node[right=4pt] at (0,.48) {$p_{12}$};
\draw[chfour external] (c)--(-.8,1.7) node[above left=2pt] {$1$};
\draw[chfour external] (c)--(.8,1.7) node[above right=2pt] {$2$};
\draw[chfour external] (b)--(0,-2.35) node[below=2pt] {$3$};
\foreach \v in {a,b,c} \node[chfour vertex] at (\v) {};
\draw[chfour momentum] (-1.05,-.5)--(-1.05,-1.1) node[midway,left=3pt] {$\ell$};
\draw[chfour momentum] (1.05,-1.1)--(1.05,-.5) node[midway,right=3pt] {$\ell+p_{12}$};
\node at (0,-2.95) {$\mathrm{NLSM}$};
\end{scope}
\end{tikzpicture}
\caption{Bubble diagrams $\Gamma_{12|3}$ with cubic vertices only. The left diagram contains a YM gluon loop. In the right diagram, the solid loop lines are $Y$ propagators and the external dashed lines are $Z$ legs. The internal dashed line carrying $p_{12}$ is a mixed $XZ$ propagator, with its $X$ end at the tree vertex and its $Z$ end at the loop vertex.}
\label{fig:ch4r-bubble}
\end{figure}

\paragraph{Bubble with cubic vertices only.}
The integrand contributions of the two diagrams in figure~\ref{fig:ch4r-bubble} are
\begin{equation}
\begin{aligned}
I_{\mathrm{YM}}^{\mathrm{bub}}
={}&\frac{1}{\ell^2(\ell+p_{12})^2}\\
&\times\Big[-\big((2\ell+2p_{12}+p_3)\cdot\varepsilon_3\big)\delta^{\mu}{}_{\nu}+\varepsilon_3^{\mu}(\ell+p_{12}+2p_3)_{\nu}
 +(\ell+p_{12}-p_3)^{\mu}\varepsilon_{3\nu}\Big]\\
&\times\Big[-\big((2\ell+p_{12})\cdot A_{12}\big)\delta^{\nu}{}_{\mu}+A_{12}^{\nu}(\ell+2p_{12})_{\mu}
 +(\ell-p_{12})^{\nu}A_{12\mu}\Big].
\end{aligned}
\label{eq:ch4-bubble-YM}
\end{equation}
\begin{equation}
I_{\mathrm{NLSM}}^{\mathrm{bub}}
=\frac{4(\ell\cdot Z_{12})[(\ell+p_{12})\cdot p_3]}
 {\ell^2(\ell+p_{12})^2}.
\label{eq:ch4-bubble-NLSM}
\end{equation}
Applying the differential operator $O$ to $I_{\mathrm{YM}}^{\mathrm{bub}}$ and simplifying the Lorentz indices gives
\begin{equation}
O I_{\mathrm{YM}}^{\mathrm{bub}}
=\frac{4(\ell\cdot Z_{12})[(\ell+p_{12})\cdot p_3]}
 {\ell^2(\ell+p_{12})^2}
=I_{\mathrm{NLSM}}^{\mathrm{bub}},
\label{eq:ch4-bubble-relation}
\end{equation}
which matches the result obtained from a direct evaluation of the NLSM bubble diagram.
\begin{figure}[H]
\centering
\begin{tikzpicture}[chfour figure]
\begin{scope}

\coordinate (a) at (-.95,0);
\coordinate (b) at (.95,0);
\coordinate (c) at (0,-1.65);
\draw[chfour gluon] (a)--(b);
\draw[chfour gluon] (b)--(c);
\draw[chfour gluon] (c)--(a);
\draw[chfour gluon] (a)--(-1.72,.6) node[above left=2pt] {$2$};
\draw[chfour gluon] (b)--(1.72,.6) node[above right=2pt] {$1$};
\draw[chfour gluon] (c)--(0,-2.4) node[below=2pt] {$3$};
\foreach \v in {a,b,c} \node[chfour vertex] at (\v) {};
\draw[chfour momentum] (-.4,.28)--(.4,.28) node[midway,above=2pt] {$\ell+p_1$};
\draw[chfour momentum] (.88,-.5)--(.55,-1.08) node[midway,right=3pt] {$\ell$};
\draw[chfour momentum] (-.55,-1.08)--(-.88,-.5) node[midway,left=3pt] {$\ell+p_{12}$};
\node at (0,-2.98) {$\mathrm{YM}$};
\end{scope}
\begin{scope}[xshift=6.0cm]

\coordinate (a) at (-.95,0);
\coordinate (b) at (.95,0);
\coordinate (c) at (0,-1.65);
\draw[chfour scalar] (a)--(b);
\draw[chfour scalar] (b)--(c);
\draw[chfour scalar] (c)--(a);
\draw[chfour external] (a)--(-1.72,.6) node[above left=2pt] {$2$};
\draw[chfour external] (b)--(1.72,.6) node[above right=2pt] {$1$};
\draw[chfour external] (c)--(0,-2.4) node[below=2pt] {$3$};
\foreach \v in {a,b,c} \node[chfour vertex] at (\v) {};
\draw[chfour momentum] (-.4,.28)--(.4,.28) node[midway,above=2pt] {$\ell+p_1$};
\draw[chfour momentum] (.88,-.5)--(.55,-1.08) node[midway,right=3pt] {$\ell$};
\draw[chfour momentum] (-.55,-1.08)--(-.88,-.5) node[midway,left=3pt] {$\ell+p_{12}$};
\node at (0,-2.98) {$\mathrm{NLSM}$};
\end{scope}
\end{tikzpicture}
\caption{One-loop three-point triangle diagrams. Wavy lines denote gluons, solid lines denote $Y$, and dashed lines denote external $Z$ legs. The internal-line momenta in both diagrams are $\ell$, $\ell+p_1$, and $\ell+p_{12}$.}
\label{fig:ch4r-triangle}
\end{figure}
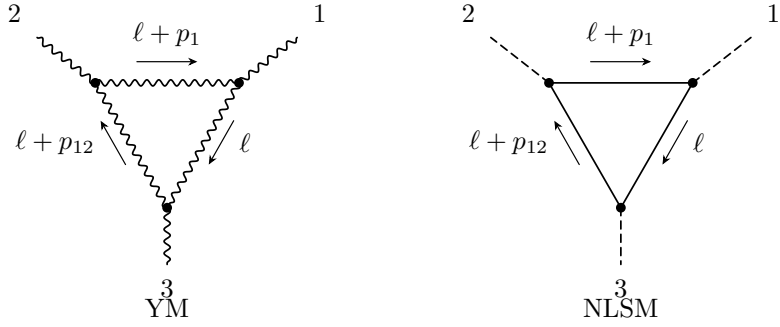

\paragraph{Triangle diagram.}
Finally, consider the triangle diagram in figure~\ref{fig:ch4r-triangle}.

The integrand contributions of the two diagrams are
\begin{equation}
\begin{aligned}
I_{\triangle}^{\mathrm{YM}}
={}&\frac{1}{\ell^2(\ell+p_1)^2(\ell+p_{12})^2}\\
&\times\Big[-\big((2\ell+2p_{12}+p_3)\cdot\varepsilon_3\big)\delta^{\mu}{}_{\nu}+\varepsilon_3^{\mu}(\ell+p_{12}+2p_3)_{\nu}
 +(\ell+p_{12}-p_3)^{\mu}\varepsilon_{3\nu}\Big]\\
&\times\Big[-\big((2\ell+2p_1+p_2)\cdot\varepsilon_2\big)\delta^{\nu}{}_{\rho}+\varepsilon_2^{\nu}(\ell+p_1+2p_2)_{\rho}
 +(\ell+p_1-p_2)^{\nu}\varepsilon_{2\rho}\Big]\\
&\times\Big[-\big((2\ell+p_1)\cdot\varepsilon_1\big)\delta^{\rho}{}_{\mu}+\varepsilon_1^{\rho}(\ell+2p_1)_{\mu}
 +(\ell-p_1)^{\rho}\varepsilon_{1\mu}\Big].
\end{aligned}
\label{eq:ch4-triangle-YM}
\end{equation}
\begin{equation}
I_{\triangle}^{\mathrm{NLSM}}
=-\frac{8(\ell\cdot p_1)[(\ell+p_1)\cdot p_2][(\ell+p_{12})\cdot p_3]}
 {\ell^2(\ell+p_1)^2(\ell+p_{12})^2}.
\label{eq:ch4-triangle-NLSM}
\end{equation}
Finally, acting with the differential operator $O$ on $I_{\triangle}^{\mathrm{YM}}$ yields
\begin{equation}
\begin{aligned}
O I_{\triangle}^{\mathrm{YM}}
&=-\frac{8(\ell\cdot p_1)[(\ell+p_1)\cdot p_2][(\ell+p_{12})\cdot p_3]}
 {\ell^2(\ell+p_1)^2(\ell+p_{12})^2}\\
&=I_{\triangle}^{\mathrm{NLSM}},
\end{aligned}
\label{eq:ch4-triangle-relation}
\end{equation}
in exact agreement with the direct calculation of the NLSM triangle diagram.

In these three-point one-loop examples, contributions containing a four-point vertex vanish after transmutation, while the integrands of the remaining diagrams correspond to those of the NLSM diagrams with the same external-leg ordering and connectivity. Although the tadpole diagram satisfies this relation, it does not contribute to the full amplitude. Thus, the diagram-by-diagram relation established at tree level by comparing propagator structures also extends, through sewing, to the one-loop integrands considered here.

\raggedbottom
\section{Conclusions and outlook}
\label{sec:conclusion}
Starting from multiparticle currents, we have studied tree-level and one-loop unifying relations between YM theory and the NLSM. Rather than acting directly on on-shell amplitudes, we allow more than one external leg to be off shell, so that trace and longitudinal operators can act directly on the recursion relations and the results can be compared term by term with the corresponding currents in the cubic $X,Y,Z$ representation of the NLSM. In this way, the unifying relations are realized not only in the final amplitudes but also at the level of the recursive structure.

At tree level, different choices of the two $Y$ legs give two types of multiparticle-current unifying relations, corresponding to the NLSM $Y$- and $X$-currents. The relation in section~\ref*{sec:x-current-relation} contains an off-shell correction $\Gamma_M$ determined by the recursion~\eqref{eq:Gamma-recursion}. This correction is generally nonzero for off-shell kinematics, but its kinematic coefficient vanishes in the amplitude limit, so it no longer contributes and the standard tree-level YM--NLSM unifying relation is recovered.

At one loop, the tree-level unifying relation extends to one-loop integrands through sewing. After sewing, the tree-level trace operator becomes a derivative with respect to the spacetime dimension $d$, while the longitudinal operator becomes $\cL_r^{\ell;M\setminus\{i,j\}}$. After excluding diagrams annihilated by the transmutation operator, the remaining YM diagrams correspond diagram by diagram to the NLSM diagrams, yielding the one-loop unifying relation.

The one-loop results of this paper provide new clues for studying higher-loop unifying relations. In recent years, a series of advances has been made in constructing multiloop integrands from classical equations of motion and off-shell recursions~\cite{Gomez:2022dzk,Lee:2022aiu,Tao:2025fch,Tao:2025biv,Tao:2026irg}. In particular, ref.~\cite{Cao:2026nlsm} presents a local quartic two-derivative representation of the NLSM capable of describing planar amplitudes at arbitrary loop order. Establishing off-shell unifying relations analogous to those in this paper within these higher-loop constructions is a question for further study.

Off-shell unifying relations also merit extension to curved spacetime. Berends--Giele unifying relations in (A)dS and analogous structures in cosmological correlators have received initial study~\cite{Tao:2022AdSUnifying,Gomez:2021CosmoSE,Armstrong:2022EFTCosmo,Chen:2023CosmoUnifying,Chen:2023dcx}. One can therefore further investigate the corresponding form of the YM--NLSM unifying relations on (A)dS backgrounds and how these relations are reflected in cosmological correlators. Introducing sewing into this framework may also lead to relations between loop-level correlators.

YM--NLSM is only one example within a larger web of unifying relations. Effective theories such as Born--Infeld theory, the NLSM, and the special Galileon have known transmutation and unifying relations at both tree level and one loop~\cite{Cachazo:2014xea,Cheung:2017ems,Zhou:2021gnu,Zhou:2022orv,Dong:2021Universal}. Thus, BI$\rightarrow$NLSM and BI$\rightarrow$SG are two natural examples to consider next. If these relations can likewise be lifted off shell and preserved under sewing, the construction presented here can be extended to a broader class of effective field theories.

\section*{Acknowledgments}
	
We would particularly like to thank Yi-Xiao Tao for invaluable discussions at every stage of this research.

\appendix
\section{Appendices}

\subsection{Proof of lemma~\ref{lem:two-y-current}}\label{eq:two-y-current-lemma-proof}
We proceed by induction on the word length \(|M|\), specifically on \(|Q|\) and \(|R|\) separately. When \(|M|=2\), necessarily \(Q=\{i\}\) and \(R=\{j\}\), providing a common base case for the induction. The single-particle boundary conditions
\[
A_i=\epsilon_i,\qquad A_j=\epsilon_j,\qquad Y_i=Y_j=1,
\]
give
\[
\cL_{M\setminus\{i,j\}}^{M}\cT_{ij}(A_i\cdot A_j)
=
\cT_{ij}(\epsilon_i\cdot \epsilon_j)
=
1
=
Y_iY_j.
\]
so lemma~\ref{lem:two-y-current} holds for \(|M|=2\).

Assume that lemma~\ref{lem:two-y-current} holds for every word of length less than \(|M|\). For \(|M|>2\), at least one of \(|Q|>1\) and \(|R|>1\) holds. First consider \(|Q|>1\). The YM multiparticle-current recursion~\eqref{eq:ym-bg-gluon}, after contracting \(A_Q\) with \(A_R\), gives
\begin{equation}
\begin{aligned}
A_Q\cdot A_R
=
\frac{1}{2s_Q}
\sum_{Q=ST}
\Bigg[
&\underbrace{
2(A_S\cdot p_T)(A_T\cdot A_R)
}_{\mathrm{(I)}}
+
\underbrace{
(p_S\cdot A_S)(A_T\cdot A_R)
}_{\mathrm{(II)}}
-
\underbrace{
(A_S\cdot A_T)(p_T\cdot A_R)
}_{\mathrm{(III)}}
\\
&-
\sum_{T=UV}
\Big(
\underbrace{
(A_S\cdot A_U)(A_V\cdot A_R)
}_{\mathrm{(IV)}}
-(U\leftrightarrow V)
\Big)
-(S\leftrightarrow T)
\Bigg].
\end{aligned}
\label{eq:two-y-bg-contracted}
\end{equation}
Apply \(\cL_{M\setminus\{i,j\}}^{M}\cT_{ij}\) to both sides of eq.~\eqref{eq:two-y-bg-contracted}. We analyze \(\mathrm{(I)}\)--\(\mathrm{(IV)}\) separately.

For \(\mathrm{(II)}\), if \(i\notin T\), then \(A_T\cdot A_R\) contains no \(\epsilon_i\), and \(\cT_{ij}\) directly gives zero. If \(i\in T\), then \(i\notin S\), and the distribution rule for operator products gives the factorization
\[
\begin{aligned}
\cL_{M\setminus\{i,j\}}^{M}\cT_{ij}
\left[(p_S\cdot A_S)(A_T\cdot A_R)\right]
=
\cL_S^{S}(p_S\cdot A_S)\,
\cL_{T\cup R\setminus\{i,j\}}^{T\cup R}
\cT_{ij}(A_T\cdot A_R)
=0,
\end{aligned}
\]
The last step uses eq.~\eqref{eq:ampm}. Thus, \(\mathrm{(II)}\) gives no contribution.

For \(\mathrm{(III)}\), \(A_S\cdot A_T\) contains no \(\epsilon_j\), while \(p_T\cdot A_R\) contains no \(\epsilon_i\), so no contraction of the form \(\epsilon_i\cdot \epsilon_j\) occurs. Hence,
\[
\cT_{ij}\left[(A_S\cdot A_T)(p_T\cdot A_R)\right]=0.
\]
so \(\mathrm{(III)}\) gives no contribution.

For \(\mathrm{(IV)}\), if \(i\notin V\), then \(A_V\cdot A_R\) contains no \(\epsilon_i\), and \(\cT_{ij}\) directly gives zero. If \(i\in V\), then \(i\notin S\cup U\), and
\[
\begin{aligned}
\cL_{M\setminus\{i,j\}}^{M}\cT_{ij}
\left[(A_S\cdot A_U)(A_V\cdot A_R)\right]
=
\cL_{S\cup U}^{S\cup U}(A_S\cdot A_U)\,
\cL_{V\cup R\setminus\{i,j\}}^{V\cup R}
\cT_{ij}(A_V\cdot A_R)
=0,
\end{aligned}
\]
The last step uses eq.~\eqref{eq:LM-AMAN-zero}. Its \(U\leftrightarrow V\) term is treated identically. Thus, \(\mathrm{(IV)}\) also gives no contribution. The corresponding \(S\leftrightarrow T\) terms of \(\mathrm{(II)}\)--\(\mathrm{(IV)}\) vanish for exactly the same reasons, so only \(\mathrm{(I)}\) and its \(S\leftrightarrow T\) term can give nonzero contributions.

Now consider \(\mathrm{(I)}\). If \(i\in T\), then \(i\notin S\), so \(S\) is a pure-\(Z\) subword, and
\begin{equation}
\begin{aligned}
\cL_{M\setminus\{i,j\}}^{M}\cT_{ij}
\left[(A_S\cdot p_T)(A_T\cdot A_R)\right]
&=
\cL_S^{Q}(A_S\cdot p_T)\,
\cL_{T\cup R\setminus\{i,j\}}^{T\cup R}
\cT_{ij}(A_T\cdot A_R).
\end{aligned}
\label{eq:two-y-I-T}
\end{equation}
Since \(S\cap T=\varnothing\) and all external legs in \(S\) are \(Z\)-states, the relation in section~\ref*{sec:z-current-relation} gives
\[
\cL_S^{Q}(A_S\cdot p_T)=Z_S\cdot p_T.
\]
On the other hand,
\[
|T\cup R|<|M|,
\]
with \(i\in T\) and \(j\in R\), while all external legs in \((T\cup R)\setminus\{i,j\}\) are still \(Z\)-states. The induction hypothesis therefore gives
\[
\cL_{T\cup R\setminus\{i,j\}}^{T\cup R}
\cT_{ij}(A_T\cdot A_R)
=
Y_TY_R.
\]
Hence, when \(i\in T\), the contribution of \(\mathrm{(I)}\) is
\[
(Z_S\cdot p_T)Y_TY_R.
\]

On the other hand, if \(i\in S\), the original \(\mathrm{(I)}\) term vanishes under \(\cT_{ij}\), while its \(S\leftrightarrow T\) term can contribute. In this case, \(i\notin T\), so \(T\) is a pure-\(Z\) subword, and
\[
\begin{aligned}
\cL_{M\setminus\{i,j\}}^{M}\cT_{ij}
\left[(A_T\cdot p_S)(A_S\cdot A_R)\right]
&=
\cL_T^{Q}(A_T\cdot p_S)\,
\cL_{S\cup R\setminus\{i,j\}}^{S\cup R}
\cT_{ij}(A_S\cdot A_R)
\\
&=
(Z_T\cdot p_S)Y_SY_R,
\end{aligned}
\]
Here, we have used the relation in section~\ref*{sec:z-current-relation} and the induction hypothesis, respectively. This term comes from \(-(S\leftrightarrow T)\) in eq.~\eqref{eq:two-y-bg-contracted} and therefore carries a minus sign.

Substituting \(\mathrm{(I)}\) and its exchanged term back into eq.~\eqref{eq:two-y-bg-contracted}, and using the factor of \(2\) in front of them to cancel the overall \(1/2\), gives
\begin{align}
\cL_{M\setminus\{i,j\}}^{M}\cT_{ij}(A_Q\cdot A_R)
&=
\frac{Y_R}{s_Q}
\Bigg[
\sum_{\substack{Q=ST\\ i\in T}}
(Z_S\cdot p_T)Y_T
-
\sum_{\substack{Q=ST\\ i\in S}}
(Z_T\cdot p_S)Y_S
\Bigg]
\nonumber\\
&=
-\frac{Y_R}{s_Q}
\sum_{Q=ST}
\Big[
Y_S(Z_T\cdot p_S)
-(S\leftrightarrow T)
\Big]
\nonumber\\
&=
Y_QY_R.
\label{eq:two-y-current-induction}
\end{align}
The last step is precisely the NLSM multiparticle-current recursion~\eqref{eq:recursion-Y} for the \(Y\)-current. Since \(i\) is the only \(Y\) external leg in \(Q\), any subword \(S\) or \(T\) that does not contain \(i\) is a pure-\(Z\) subword. Its corresponding \(Y_S\) or \(Y_T\) automatically vanishes, as established in section~\ref{sec:y-current-relation} by lemma~\ref{lem:even-non-z-y-zero}. Thus, the second equality includes both \(i\in S\) and \(i\in T\).

Finally, consider \(|Q|=1\). Then \(Q=\{i\}\), and since \(|M|>2\), necessarily \(|R|>1\). Using
\[
A_Q\cdot A_R=A_R\cdot A_Q,\qquad \cT_{ij}=\cT_{ji},
\]
and exchanging
\[
(Q,i)\longleftrightarrow(R,j)
\]
allows the same multiparticle-current recursion analysis to be applied to \(A_R\). Here, \(j\) in \(R\) plays the role that \(i\) in \(Q\) played above, and \(Y_Q=Y_i=1\). The same induction steps give
\[
\cL_{M\setminus\{i,j\}}^{M}\cT_{ij}(A_Q\cdot A_R)
=
Y_RY_Q
=
Y_QY_R.
\]
Thus, lemma~\ref{lem:two-y-current} holds for every \(M=QR\) satisfying the stated conditions, completing the induction.

\subsection{Proof of lemma~\ref{lem:x-current}}\label{eq:x-current-lemma-proof}
We carry out the induction on the lengths of $Q$ and $R$ separately. In particular, when performing the induction on $|R|$, we must first establish the lemma for each possible subword $N$ of $R$ whose endpoints are $i$ and $j$. With this as the base case, we then fix $N$ and proceed by induction on the length of $R$ over all words containing $N$.

As can be seen from the proof below, when $i$ and $j$ lie at the endpoints of the word, the general argument simplifies: the steps that would require the induction hypothesis for Lemma~4 itself are absent. Thus, Lemma~4 holds in this case without invoking its own induction hypothesis. To avoid repeating the lengthy argument, we omit a separate proof of this case, which the reader can verify by following the steps below.

We first consider $|R|>2$, keeping fixed the contiguous subword $N$ of $R$ whose endpoints are $i$ and $j$. Contracting the YM multiparticle-current recursion for \(A_R\) with \(A_Q\) gives
\begin{equation}
\begin{aligned}
A_R\cdot A_Q
=
\frac{1}{2s_R}
\sum_{R=ST}
\Bigg[
&\underbrace{
2(A_S\cdot p_T)(A_T\cdot A_Q)
}_{\mathrm{(I)}}
+
\underbrace{
(p_S\cdot A_S)(A_T\cdot A_Q)
}_{\mathrm{(II)}}
-
\underbrace{
(A_S\cdot A_T)(p_T\cdot A_Q)
}_{\mathrm{(III)}}
\\
&-
\sum_{T=UV}
\Big(
\underbrace{
(A_S\cdot A_U)(A_V\cdot A_Q)
}_{\mathrm{(IV)}}
-(U\leftrightarrow V)
\Big)
-(S\leftrightarrow T)
\Bigg].
\end{aligned}
\label{eq:x-current-bg}
\end{equation}
Apply \(\cL_{M\setminus\{i,j\}}^{M}\cT_{ij}\) to both sides of eq.~\eqref{eq:x-current-bg}. We analyze \(\mathrm{(I)}\)--\(\mathrm{(IV)}\) separately.

For \(\mathrm{(II)}\), \(\cT_{ij}\) can give a nonzero contribution when acting on \(A_T\cdot A_Q\) only when both labels \(i\) and \(j\) lie in \(T\). In that case, however,
\[
\begin{aligned}
\cL_{M\setminus\{i,j\}}^{M}\cT_{ij}
\left[(p_S\cdot A_S)(A_T\cdot A_Q)\right]
&=
\cL_S^S(p_S\cdot A_S)\,
\cL_{T\cup Q\setminus\{i,j\}}^{T\cup Q}
\cT_{ij}(A_T\cdot A_Q)
=0,
\end{aligned}
\]
where we have used the preceding result \(\cL_S^S(p_S\cdot A_S)=0\). Thus, \(\mathrm{(II)}\) gives no contribution. For \(\mathrm{(IV)}\), under \(\cL_{M\setminus\{i,j\}}^{M}\cT_{ij}\), the four-current product is either annihilated by $\cT_{ij}$ because $\epsilon_i$ and $\epsilon_j$ occur in different current contractions, or vanishes because the distribution rule for operator products yields an expression of the form in eq.~\eqref{eq:LM-AMAN-zero}. Thus, only \(\mathrm{(I)}\) and \(\mathrm{(III)}\) need to be analyzed.

First consider \(\mathrm{(I)}\). If \(i,j\in S\), then
\[
\cL_{M\setminus\{i,j\}}^{M}\cT_{ij}
\left[(A_S\cdot p_T)(A_T\cdot A_Q)\right]
=
\left(
\cL_{S\setminus\{i,j\}}^{R}\cT_{ij}(A_S\cdot p_T)
\right)
\left(
\cL_{T\cup Q}^{T\cup Q}(A_T\cdot A_Q)
\right)
=0.
\]
Similarly, if \(i,j\in T\), then
\[
\cL_{M\setminus\{i,j\}}^{M}\cT_{ij}
\left[(A_T\cdot p_S)(A_S\cdot A_Q)\right]
=0.
\]
Thus, in the \(i,j\in S\) sector, only the exchanged term
$
(A_T\cdot p_S)(A_S\cdot A_Q)
$
can contribute. Since \(T\) is a pure-\(Z\) subword, the relation in section~\ref*{sec:z-current-relation} and the induction hypothesis give
\[
\begin{aligned}
\cL_{M\setminus\{i,j\}}^{M}\cT_{ij}
\left[(A_T\cdot p_S)(A_S\cdot A_Q)\right]
&=
\cL_T^R(A_T\cdot p_S)\,
\cL_{Q\cup S\setminus\{i,j\}}^{Q\cup S}
\cT_{ij}(A_S\cdot A_Q)
\\
&=
(Z_T\cdot p_S)
\left(
X_S\cdot Z_Q+\frac{Z_Q\cdot p_S}{2s_S}\,\Gamma_S
\right).
\end{aligned}
\]
Similarly, in the \(i,j\in T\) sector,
\[
\begin{aligned}
\cL_{M\setminus\{i,j\}}^{M}\cT_{ij}
\left[(A_S\cdot p_T)(A_T\cdot A_Q)\right]
&=
(Z_S\cdot p_T)
\left(
X_T\cdot Z_Q+\frac{Z_Q\cdot p_T}{2s_T}\,\Gamma_T
\right).
\end{aligned}
\]
If \(i\in S\) and \(j\in T\), then \(\epsilon_i,\epsilon_j\) cannot form \(\epsilon_i\cdot \epsilon_j\) in
$
(A_S\cdot p_T)(A_T\cdot A_Q)
$
or its exchanged term, so \(\cT_{ij}\) directly annihilates both terms.

Next consider \(\mathrm{(III)}\). If \(i,j\in S\), the induction hypothesis and the relation in section~\ref*{sec:z-current-relation} give
\[
\begin{aligned}
\cL_{M\setminus\{i,j\}}^{M}\cT_{ij}
\left[(A_S\cdot A_T)(p_T\cdot A_Q)\right]
&=
\cL_{R\setminus\{i,j\}}^R
\cT_{ij}(A_S\cdot A_T)\,
\cL_Q^{Q\cup T}(p_T\cdot A_Q)
\\
&=
\left(
X_S\cdot Z_T+\frac{Z_T\cdot p_S}{2s_S}\,\Gamma_S
\right)
(Z_Q\cdot p_T),
\end{aligned}
\]
while the exchanged term gives
\[
\cL_{M\setminus\{i,j\}}^{M}\cT_{ij}
\left[(A_T\cdot A_S)(p_S\cdot A_Q)\right]
=
\left(
X_S\cdot Z_T+\frac{Z_T\cdot p_S}{2s_S}\,\Gamma_S
\right)
(Z_Q\cdot p_S).
\]
If \(i,j\in T\), we similarly obtain
\[
\begin{aligned}
\cL_{M\setminus\{i,j\}}^{M}\cT_{ij}
\left[(A_S\cdot A_T)(p_T\cdot A_Q)\right]
&=
\left(
X_T\cdot Z_S+\frac{Z_S\cdot p_T}{2s_T}\,\Gamma_T
\right)
(Z_Q\cdot p_T),
\\
\cL_{M\setminus\{i,j\}}^{M}\cT_{ij}
\left[(A_T\cdot A_S)(p_S\cdot A_Q)\right]
&=
\left(
X_T\cdot Z_S+\frac{Z_S\cdot p_T}{2s_T}\,\Gamma_T
\right)
(Z_Q\cdot p_S).
\end{aligned}
\]
Finally, if \(i\in S\) and \(j\in T\), lemma~\ref{lem:two-y-current} gives
\[
\begin{aligned}
\cL_{M\setminus\{i,j\}}^{M}\cT_{ij}
\left[(A_S\cdot A_T)(p_T\cdot A_Q)\right]
&=
Y_SY_T\,(Z_Q\cdot p_T),
\\
\cL_{M\setminus\{i,j\}}^{M}\cT_{ij}
\left[(A_T\cdot A_S)(p_S\cdot A_Q)\right]
&=
Y_SY_T\,(Z_Q\cdot p_S).
\end{aligned}
\]

Substituting all the nonzero contributions above back into eq.~\eqref{eq:x-current-bg}, and using \(p_R=p_S+p_T\) to collect all terms containing \(\Gamma\), gives
\begingroup\small
\begin{align}
&\cL_{M\setminus\{i,j\}}^{M}\cT_{ij}(A_R\cdot A_Q)
\nonumber\\
&=
\frac{1}{2s_R}
\Bigg[
\sum_{\substack{R=ST\\ i,j\in S}}
\Big(
-2Z_T\cdot p_S\,X_S\cdot Z_Q
+2X_S\cdot Z_T\,Z_Q\cdot p_S
\Big)
-\sum_{\substack{R=ST\\ i\in S,\ j\in T}}
Y_SY_T\left(Z_Q\cdot p_T-Z_Q\cdot p_S\right)
\nonumber\\
&\qquad
+\sum_{\substack{R=ST\\ i,j\in T}}
\Big(
2Z_S\cdot p_T\,X_T\cdot Z_Q
-2X_T\cdot Z_S\,Z_Q\cdot p_T
\Big)
\Bigg]
\nonumber\\
&\quad
+\frac{Z_Q\cdot p_R}{2s_R}
\Bigg[
\sum_{\substack{R=ST\\ i,j\in T}}
\left(
Z_S\cdot X_T
+\frac{Z_S\cdot p_T}{2s_T}\Gamma_T
\right)
-
\sum_{\substack{R=ST\\ i,j\in S}}
\left(
Z_T\cdot X_S
+\frac{Z_T\cdot p_S}{2s_S}\Gamma_S
\right)
\Bigg].
\label{eq:x-current-separated}
\end{align}
\endgroup
By the recursion~\eqref{eq:Gamma-recursion} for \(\Gamma_R\), the last line is precisely
$
\frac{Z_Q\cdot p_R}{2s_R}\Gamma_R.
$

We now show that the first two lines of eq.~\eqref{eq:x-current-separated} equal \(Z_Q\cdot X_R\). The NLSM multiparticle-current recursion for the \(X\)-current gives
\[
\begin{aligned}
Z_Q\cdot X_R
=
-\frac{1}{s_R}
\sum_{R=ST}
\Big[
(Z_Q\cdot p_T)
\left(
\frac12Y_SY_T+Z_S\cdot X_T
\right)
+
(p_S\cdot Z_T)(X_S\cdot Z_Q)
-(S\leftrightarrow T)
\Big].
\end{aligned}
\]
We discuss the three sectors \(i,j\in S\), \(i\in S,j\in T\), and \(i,j\in T\) separately. If \(i,j\in S\), lemmas~\ref{lem:non-z-z-zero} and~\ref{lem:even-non-z-y-zero} imply
\[
Y_SY_T=0,\qquad
Z_S\cdot X_T=0,\qquad
p_T\cdot Z_S=0,
\]
so the contribution of this sector is
\[
\frac{1}{2s_R}
\sum_{\substack{R=ST\\ i,j\in S}}
\Big(
-2Z_T\cdot p_S\,X_S\cdot Z_Q
+2X_S\cdot Z_T\,Z_Q\cdot p_S
\Big).
\]
If \(i\in S,j\in T\), the corresponding \(XZ\) terms all vanish, leaving only
\[
-\frac{1}{2s_R}
\sum_{\substack{R=ST\\ i\in S,\ j\in T}}
Y_SY_T
\left(
Z_Q\cdot p_T-Z_Q\cdot p_S
\right).
\]
If \(i,j\in T\), we similarly obtain
\[
\frac{1}{2s_R}
\sum_{\substack{R=ST\\ i,j\in T}}
\Big(
2Z_S\cdot p_T\,X_T\cdot Z_Q
-2X_T\cdot Z_S\,Z_Q\cdot p_T
\Big).
\]
The sum of these three parts is exactly the first two lines of eq.~\eqref{eq:x-current-separated}. Thus,
\[
\cL_{M\setminus\{i,j\}}^{M}\cT_{ij}(A_R\cdot A_Q)
=
Z_Q\cdot X_R+\frac{Z_Q\cdot p_R}{2s_R}\Gamma_R.
\]
so the induction step for \(|R|>2\) holds.

We must also consider the induction step for \(|Q|>1\). In this case, apply the YM multiparticle-current recursion to \(A_Q\). Since both \(i,j\) lie in \(R\), the four-current terms and the terms containing \(A_S\cdot A_T\) vanish for the same reasons as above, leaving only

\[
\begin{aligned}
\cL_{M\setminus\{i,j\}}^{M}\cT_{ij}(A_R\cdot A_Q)
=
\frac{1}{s_Q}
\sum_{Q=ST}
\Big[
\cL_S^Q(A_S\cdot p_T)\,
\cL_{T\cup R\setminus\{i,j\}}^{T\cup R}
\cT_{ij}(A_T\cdot A_R)
-(S\leftrightarrow T)
\Big].
\end{aligned}
\]

Since both \(S,T\) are pure-\(Z\) subwords, the relation in section~\ref*{sec:z-current-relation} and the induction hypothesis give
\begin{equation}
\begin{aligned}
\cL_{M\setminus\{i,j\}}^{M}\cT_{ij}(A_R\cdot A_Q)
={}&
\frac{1}{s_Q}
\sum_{Q=ST}
\Big[
(Z_S\cdot p_T)(X_R\cdot Z_T)
-(Z_T\cdot p_S)(X_R\cdot Z_S)
\Big]
\\
&+
\frac{\Gamma_R}{2s_Rs_Q}
\sum_{Q=ST}
\Big[
(Z_S\cdot p_T)(Z_T\cdot p_R)
-(Z_T\cdot p_S)(Z_S\cdot p_R)
\Big].
\end{aligned}
\label{eq:XQ-current-separated}
\end{equation}
The second line is exactly
$
\frac{Z_Q\cdot p_R}{2s_R}\Gamma_R.
$

The multiparticle-current recursion for \(Z_Q\),
\[
Z_Q^\mu
=
-\frac{1}{s_Q}
\sum_{Q=ST}
\left[
Z_S^\mu(p_S\cdot Z_T)
-
Z_T^\mu(p_T\cdot Z_S)
\right],
\]
shows that the first line of eq.~\eqref{eq:XQ-current-separated} is exactly \(Z_Q\cdot X_R\). Thus,
\[
\cL_{M\setminus\{i,j\}}^{M}\cT_{ij}(A_R\cdot A_Q)
=
Z_Q\cdot X_R+\frac{Z_Q\cdot p_R}{2s_R}\Gamma_R.
\]
Both induction steps now hold, so lemma~\ref{lem:x-current} is valid for every word satisfying \(M=QR\) and \(i,j\in R\). This proves the statement.

\bibliographystyle{JHEP}
\bibliography{references}

\end{document}